\documentclass{article}
\usepackage[utf8]{inputenc}
\usepackage{url,hyperref,lineno,microtype}
\usepackage[onehalfspacing]{setspace}
\usepackage{natbib}     
\setcitestyle{numbers}
\usepackage{graphicx}
\usepackage{tabularx}
\usepackage{amsmath}
\usepackage{amssymb}

\usepackage{blindtext}

\usepackage{chemformula}
\usepackage{siunitx}

\usepackage[utf8]{inputenc}
\usepackage{esvect}
\usepackage{amsmath}
\usepackage{amssymb}
\usepackage[onehalfspacing]{setspace}
\usepackage{graphicx}
\usepackage{float}
\usepackage{upgreek}
\usepackage{hyperref}
\usepackage{textcomp}
\usepackage{tabularx}
\usepackage{textcomp}

\usepackage{enumitem}

\graphicspath{{img/}}

\usepackage{geometry}
\begin{document}

\begin{flushleft}
{\Large
\textbf\newline{Quantification of metabolic niche occupancy dynamics in a Baltic Sea bacterial community}
}
\newline
\\
Jana C. Massing\textsuperscript{1,2,3*},
Ashkaan Fahimipour\textsuperscript{4},
Carina Bunse\textsuperscript{1,5},
Jarone Pinhassi\textsuperscript{6},
Thilo Gross\textsuperscript{1,2,3}
\\
\bigskip
{1} Helmholtz Institute for Functional Marine Biodiversity at the University of Oldenburg (HIFMB), Oldenburg, Germany
\\
{2}  Helmholtz Centre for Marine and Polar Research, Alfred-Wegener-Institute, Bremerhaven, Germany
\\
{3} Institute for Chemistry and Biology of the Marine Environment (ICBM), Carl-von-Ossietzky University, Oldenburg, Germany
\\
{4} Institute of Marine Sciences, University of California, Santa Cruz, CA, USA
\\
{5} Department of Marine Sciences, University of Gothenburg, Gothenburg, Sweden 
\\
{6} Centre for Ecology and Evolution in Microbial Model Systems - EEMiS, Linnaeus University, Kalmar, Sweden
\\
\bigskip
* jana.massing@hifmb.de

\end{flushleft}

\section*{Abstract}
Progress in molecular methods has enabled the monitoring of bacterial populations in time. Nevertheless, understanding community dynamics and its links with ecosystem functioning remains challenging due to the tremendous diversity of microorganisms. Conceptual frameworks that make sense of time-series of taxonomically-rich bacterial communities, regarding their potential ecological function, are needed. A key concept for organizing ecological functions is the niche, the set of strategies that enable a population to persist and define its impacts on the surroundings. Here we present a framework based on manifold learning, to organize genomic information into potentially occupied bacterial metabolic niches over time. We apply the method to re-construct the dynamics of putatively occupied metabolic niches using a long-term bacterial time-series from the Baltic Sea, the Linnaeus Microbial Observatory (LMO).
The results reveal a relatively low-dimensional space of occupied metabolic niches comprising groups of taxa with similar functional capabilities.
Time patterns of occupied niches were strongly driven by seasonality. Some metabolic niches were 
dominated by one bacterial taxon whereas others were occupied by multiple taxa, and this depended on season. These results illustrate the power of manifold learning approaches to advance our understanding of the links between community composition and 
functioning in microbial systems.

\section*{Importance}
The increase in data availability of bacterial communities highlights the need for conceptual frameworks to advance our understanding of these complex and diverse communities alongside the production of such data. To understand the dynamics of these tremendously diverse communities, we need tools to identify overarching strategies and describe their role and function in the ecosystem in a comprehensive way. Here, we show that a manifold learning approach can coarse grain bacterial communities in terms of their metabolic strategies and that we can thereby quantitatively organize genomic information in terms of potentially occupied niches over time. This approach therefore advances our understanding of how fluctuations in bacterial abundances and species composition can relate to ecosystem functions and it can facilitate the analysis, monitoring and future predictions of the development of microbial communities.

\section*{Keywords}niche, marine, manifold learning, diffusion map, bacterial communities

\section{Introduction}

More than 60 years ago, Hutchinson formulated the paradox of plankton, expressing 
astonishment at the enormous diversity of organisms in the face of an apparently limited number of resources \cite{hutchinson1961paradox}. Today, the estimated diversity of 
microbial species in the ocean
extends this apparent contradiction 
to marine bacterial communities \cite{locey2016scaling, louca2019census}. Over 40,000 marine microbial species have been detected so far \cite{sunagawa2015structure, zhang2019marine}, and these microorganisms are critical for life in the oceans and on land because of 
their capacity to perform sophisticated and diverse chemical reactions, to drive major biogeochemical cycles \cite{falkowski2008microbial}, and to exhibit diverse interactions among each other and with macroorganisms \cite[e.g.][]{hadfield2011biofilms, bourne2016insights}. In addition to their huge diversity overall, marine bacterial communities undergo complex dynamic fluctuations in abundances and species compositions on the daily, monthly and annual scale \cite{bunse2017marine, fuhrman2015marine}. In surface waters in temperate and polar regions, bacterial communities show strong seasonal patterns driven by changes in multiple interacting environmental features \cite{bunse2017marine}. 

Due to the recent advances in modern molecular methods it has become possible to monitor bacterial community composition over time and establish short and long-term time series \cite[examples in][]{buttigieg2018marine}. These revealed for instance stability in average community composition in spite of strong variation on shorter scales \cite{fuhrman2015marine}, environmental selection as an important driver of seasonal community succession \cite{auladell2021seasonal} and the significance of biological interactions among bacteria themselves, and between bacteria and other organisms e.g. phytoplankton, in shorter scale community dynamics \cite{needham2016pronounced}. 

Still, the tremendous diversity of bacteria poses a challenge for data analysis. Suppose for example that we sample a bacterial community and record the relative population densities for each detected taxon \textsl{e.g.,} amplicon sequence variants, operational taxonomic units, species).
In this case, the number of variables per sample is identical to the number of taxa. Mathematically we can say that the dimensionality of the data space equals the number of taxa. Dealing with such high-dimensional data is inherently difficult (the so-called curse of dimensionality) \cite{bellman1957dynamic}. In particular in high-dimensional spaces, comparisons between all but the most similar data points become so noisy that they hurt rather than help the analysis \cite{moon2019visualizing}. 
Hence analysis can benefit from a coarse-graining step in which the dimensionality of the data set is reduced to a smaller number of variables. If done well this reduction yields new informative variables and also greatly reduces the noise in the data \cite{barter2019manifold, ghafourian2020wireless}.

When dealing with high-dimensional data spaces only a tiny portion of space is typically inhabited by data points. These data may approximately trace a curve, a curved surface, or some other comparatively low-dimensional object within the data space. 
Such objects in data-space are called data manifolds, and the task of locating them is known as manifold learning. Because the dimensionality of the manifold is lower than that of the embedding data space, manifold learning allows us to reduce the complexity of the data without losing information \cite{ma2012manifold}. 

A widely used de-facto manifold learning method is principal component analysis (PCA) \cite{jolliffe2016principal}. PCA is a linear method that can only approximate the curved manifolds in the data by a flat surface. Therefore, it often works well in identifying global contrasts and for linearly distributed data. However, it cannot handle complex nonlinear data adequately. 
An alternative is offered by diffusion maps \cite{coifman2005geometric,  coifman2006diffusion, fahimipour2020mapping}. The diffusion map identifies pairs of data points that are similar enough to be trusted even in the high dimensional space. Comparisons between any data points can then be made safely by measuring their distance on the network of all trusted comparisons. Mathematically the diffusion map is of similar complexity as PCA. However, unlike PCA it can identify curved manifolds and is immune from the curse of dimensionality.
Like PCA, diffusion maps identify a new set of variables that describe the data in a coordinate system that follows the manifolds. Recent papers demonstrate that the application of diffusion maps to ecological data \cite{fahimipour2020mapping, ryabov2021trait, reiter2021charting} yields new variables that can be interpreted as composite functional strategies. The diffusion map thus relates to the fundamental ecological niche space in a system \cite{fahimipour2020mapping, hutchinson1957cold}.

Here we apply diffusion maps to re-construct the metabolic niche space of a bacterial community from a long-term time-series in the Baltic Sea. We use the newly predicted variables to convert the taxonomic time-series into multiple strategy time-series and to quantify changes in functional diversity. Our results indicate that the diffusion map can reveal interpretable ecological strategies in the Baltic Sea bacterial community. 
This provides a quantitative framework to organize genomic information into potentially occupied ecological niches over time.

\section{Results and Discussion}
Our goal is to describe communities in terms of a coarse-grained functional coordinate system. Using diffusion maps, the coordinate axes that span this functional space can be found by a simple mathematical procedure (see Methods). The identified functional axes are new variables that are composites of functional capabilities of the analyzed bacterial community. These new variables are ordered according to their importance in structuring the dataset. Hence, in the following we refer to variable 1 as the most important, variable 2 the second most important and so on.

Moreover, the diffusion map assigns each species a score on these functional axes, i.e. a score for each new variable. Following Fahimipour and Gross \cite{fahimipour2020mapping} we interpret these new variables as effective composite metabolic strategies of the taxa. As each taxon from the dataset is assigned a score for each new variable, we can use these scores to order taxa from the most negative to the most positive entries for each variable. By analyzing the functional capabilities of exemplar taxa near the axis extrema, we can interpret the metabolic strategies that are described by the newly identified variables. For the interpretation, we divide each variable into positive and negative side because the taxa at the two sides can be characterized by different composite metabolic strategies.   

\begin{figure} [H]
    \centering
    \includegraphics[width=\textwidth]{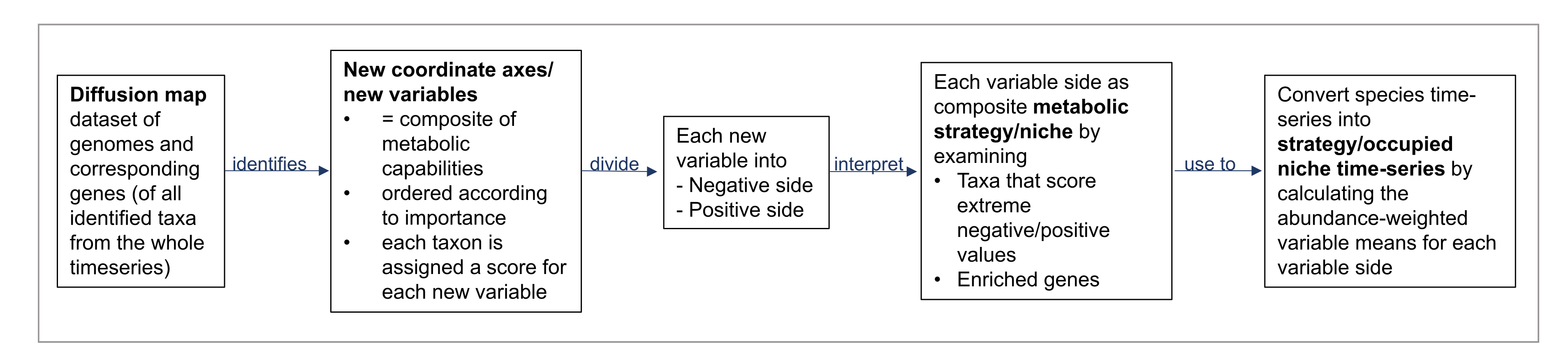}
    \caption{Overview of the procedure from diffusion mapping the dataset of genomes and genes to conversion of the species time-series into strategy time-series.}
    \label{method_plot}
\end{figure}

Variable scores (from negative and positive side of the variable) can be used to convert the taxonomic time-series into multiple strategy time-series by calculating the abundance-weighted means of each variable side for each time-point (Figure~\ref{method_plot}). This enables us to observe the dynamics of community composition over time in terms of the dynamics of putatively occupied metabolic strategies, i.e. niches. Altogether the newly identified variables span the metabolic niche space of the analyzed community.

\subsection{Important metabolic strategies.}
\begin{figure} [H]
    \centering
    \includegraphics[width=\textwidth]{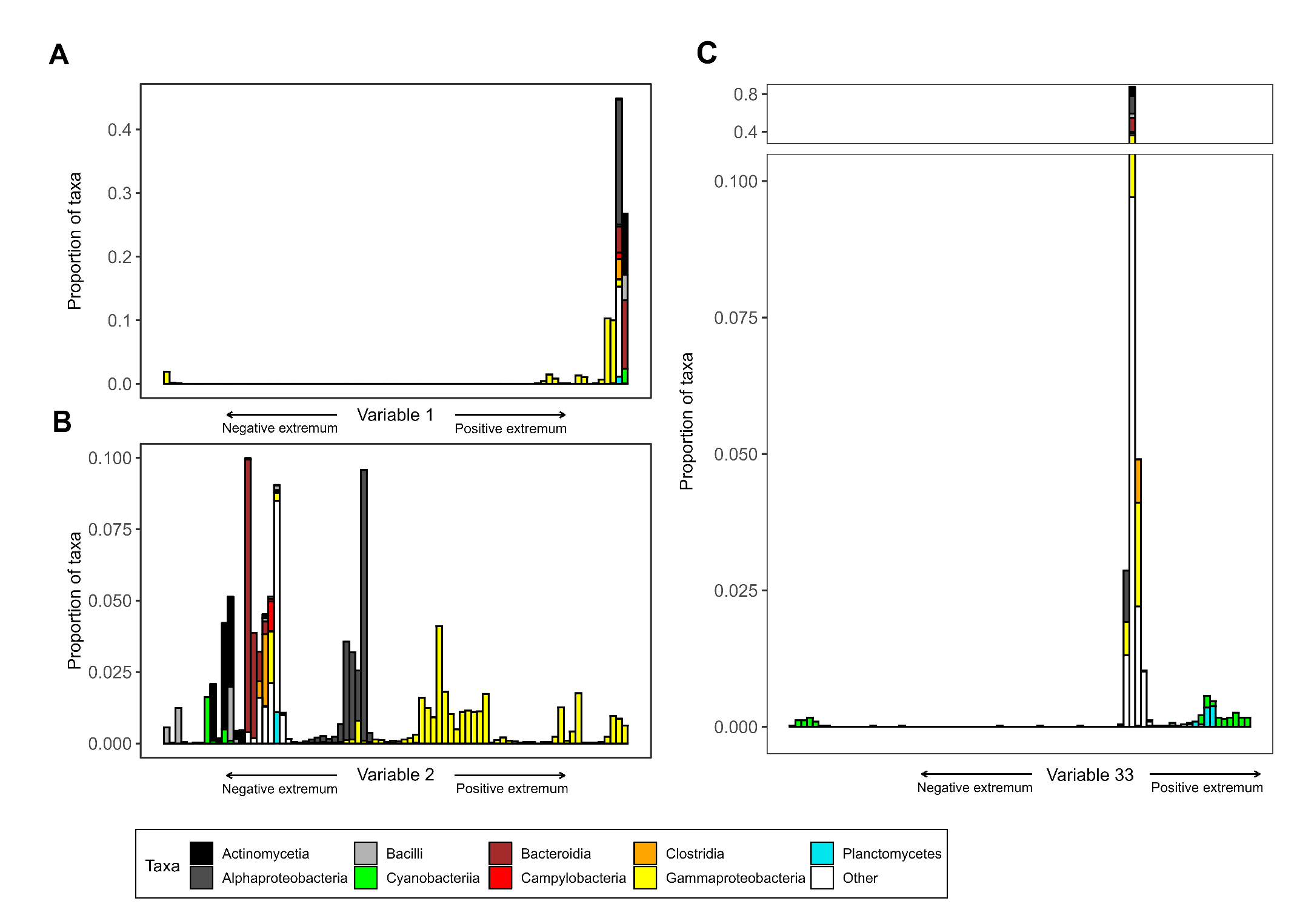}
    \caption{The ordering of taxa defined by variable 1 (A), variable 2 (B) and variable 33 (C) entries, from negative to positive (left to right). The taxonomic compositions corresponding to variable entries are shown for each of 80 equally spaced bins. }
    \label{histo_plot}
\end{figure}

The most important variable identified via diffusion mapping, variable 1, separates primarily animal- or human-associated members of the Enterobacteriaceae \emph{e.g.} close relatives of \emph{Salmonella}, \emph{Cronobacter} and \emph{Yersinia} from all other taxa (Supplementary Table~2). A collection of 92 genomes from representatives of the Enterobacteriaceae score low values, whereas all other genomes are assigned values near zero (Figure~\ref{histo_plot}~A). Such a variable in which the majority of species score close-to-zero is said to be a localized variable of the diffusion map \cite{nyberg2015mesoscopic}, because it appears as a result of Anderson localization \cite{anderson1958absence}. In diffusion map results, the appearance of localized variables identifies a clear cluster of species that is well separated from the rest. 

To reveal which metabolic capabilities characterize these taxa and distinguish them from all others, we identified genes that were over-represented in the genomes of the taxa scoring far-from-zero entries (see \textsl{Methods}). Genes encoding machinery for iron acquisition common in Enterobacteriaceae, such as the Enterobactin synthase component F \cite{pollack1970enterobactin, o1970structure}, genes responsible for the flagellar formation \cite{aldridge2002regulation} and genes associated to biofilm formation \cite{domka2006ylih} are enriched in these taxa (Supplementary Table~3). 

Despite very low abundances of the respective taxa in the samples (mean relative abundances of 0.007 over all samples, see also SI material for abundance data), this variable appears first in the diffusion map. This may at first appear a surprising result given what is known about the ecological role of Enterobacteraceae in marine ecosystems. However, it makes sense that this group is so clearly separated from the other bacterial taxa due to well-known biases toward sequences of pathogenic taxa and genes involved in pathogenesis in global databases \cite[e.g.][]{blackwell2021exploring}. Consequently, these taxa are characterized as very different to other bacterial species in the marine community. 
Because the 'full' size fraction was sampled, we might have captured plankton pathogens (e.g. from zooplankton microbiomes or similar) that are very close relatives to these Enterobacteriaceae separated by the first diffusion variable. 
Hence, this variable separates the most different bacterial taxa in terms of their known gene composition from the rest of the community, demonstrating the power of the diffusion map method to reveal such differences and to identify biases in the dataset.  

The diffusion map reveals further localized variables that represent relevant metabolic strategies for the Baltic Sea bacterial community, \emph{e.g.} variable 4 negative, which separates the Cyanobacteria from all other taxa (Supplementary Figure~3). The cyanobacterial genomes are assigned large negative values, whereas all other taxa score positive or close-to-zero values. Genes that encode the subunits of photosystem I and photosystem II as well as associated cytochrome components and cyanobacterial-specific light-harvesting antennae \cite{hallenbeck2017modern} are among the enriched genes, indicating that this variable detects cyanobacterial photosynthesis (Supplementary Table~6). Supporting the findings of a previous study \cite{fahimipour2020mapping}, the localized character of this variable reveals that cyanobacterial photosynthesis is a yes-or-no strategy, indicating that this photosynthetic lifestyle has wide-ranging metabolic consequences with distinct implications. For example, oxygenic photosynthetic lifestyle is expensive in terms of avoiding or repairing photoinhibition and -damage therefore many costly adaptations are necessary and the energy spent can not be invested into other metabolic pathways \cite{raven2011cost}.  

There are also variables that span a continuum of strategies, such as variable 2 and 3. In variable 2, marine host-associated Gammaproteobacteria, e.g. \emph{Vibrio}, \emph{Shewanella} and \emph{Photobacterium}, are found at the positive extremum, whereas oligotrophic Gamma- and Alphaproteobacteria are assigned values close to zero (Figure~\ref{histo_plot}~B). Among the most correlated capabilities for the taxa at the positive end of variable 2 are chemotaxis and response to various stressors (Supplementary Table~4). Variable 3 identifies the different strategies of marine Alphaproteobacteria: We find the Rhodobacteraceae and the Rhizobiales, known for their capability of utilizing a variety of carbon sources \cite{luo2015divergent, alves2014family}, at the positive extremum (Supplementary Figure~2). Major enriched genes encode machinery for the utilization of various dissolved organic carbon compounds (Supplementary Table~5), \emph{e.g.} phosphonate, acetate and urea that constitute exudates of phytoplankton \cite{zecher2020evidence}. The streamlined genomes of the free-living Pelagibacterales and the obligate intracellular pathogens Rickettsiales score close-to-zero values. Hence we interpret variable 2 positive as metabolic strategy of marine host-associated Gammaproteobacteria and variable 3 positive as metabolic strategy of Alphaproteobacteria able of using a wide range of carbon sources. 

Taxa that are grouped together in one variable can be separated by another variable. For example, Cyanobacteria group together in variable 4 negative, whereas they are split in variable 33: The Picocyanobacteria score extreme negative values, whereas the other cyanobacterial genomes group towards the positive side, with the heterocyst-forming family Nostocaceae scoring highest values (Figure~\ref{histo_plot}~C and Supplementary Figure~6). The Enterobacterales that score highest in variable 2 are separated in variable 27, for which the family Shewanellaceae scores extreme negative values and the other enterobacterial families, \emph{e.g.} the Vibrionaceae score positive values (Supplementary Figure~5). Variable 14 separates the Bacteroidota into anaerobic, intestinal Bacteroidota e.g. \emph{Prevotella} \cite{tett2021prevotella} on the positive side and complex polysaccharide degraders e.g. \textsl{Flavobacterium} \cite{buchan2014master} on the negative side (Supplementary Figure~4). Enriched genes for the latter encode different CAZymes \cite{teeling2012substrate}, responsible for the degradation of major plant cell wall components (Supplementary Table~7).    

Diffusion mapping also identified strategies that group genera from different taxonomic groups together, we find for example bacteria that oxidize methyl groups and C1 compounds, such as methanol and formaldehyde from different families such as Beijerinckiaceae, Xanthobacteraceae, Acetobacteraceae at the negative end of variable 38 (Supplementary Figure~7). Most correlated genes (Supplementary Table~8) encode machinery for methanol and formaldehyde degradation \cite{sun2011one}. Another example is variable 43 positive that groups the non-spore forming sulfate-reducing bacteria from the families Desulfocapsaceae, Desulfobacteraceae, De\-sul\-fur\-i\-vib\-ri\-o\-na\-ce\-ae and others together (Supplementary Figure~8), enriched genes (Supplementary Table~9) are responsible for sulfate respiration \cite{venceslau2010qrc}. 
\newline

\subsection{Structure of the inferred niche space. }
\begin{figure} [H]
    \centering
    \includegraphics[width=\textwidth]{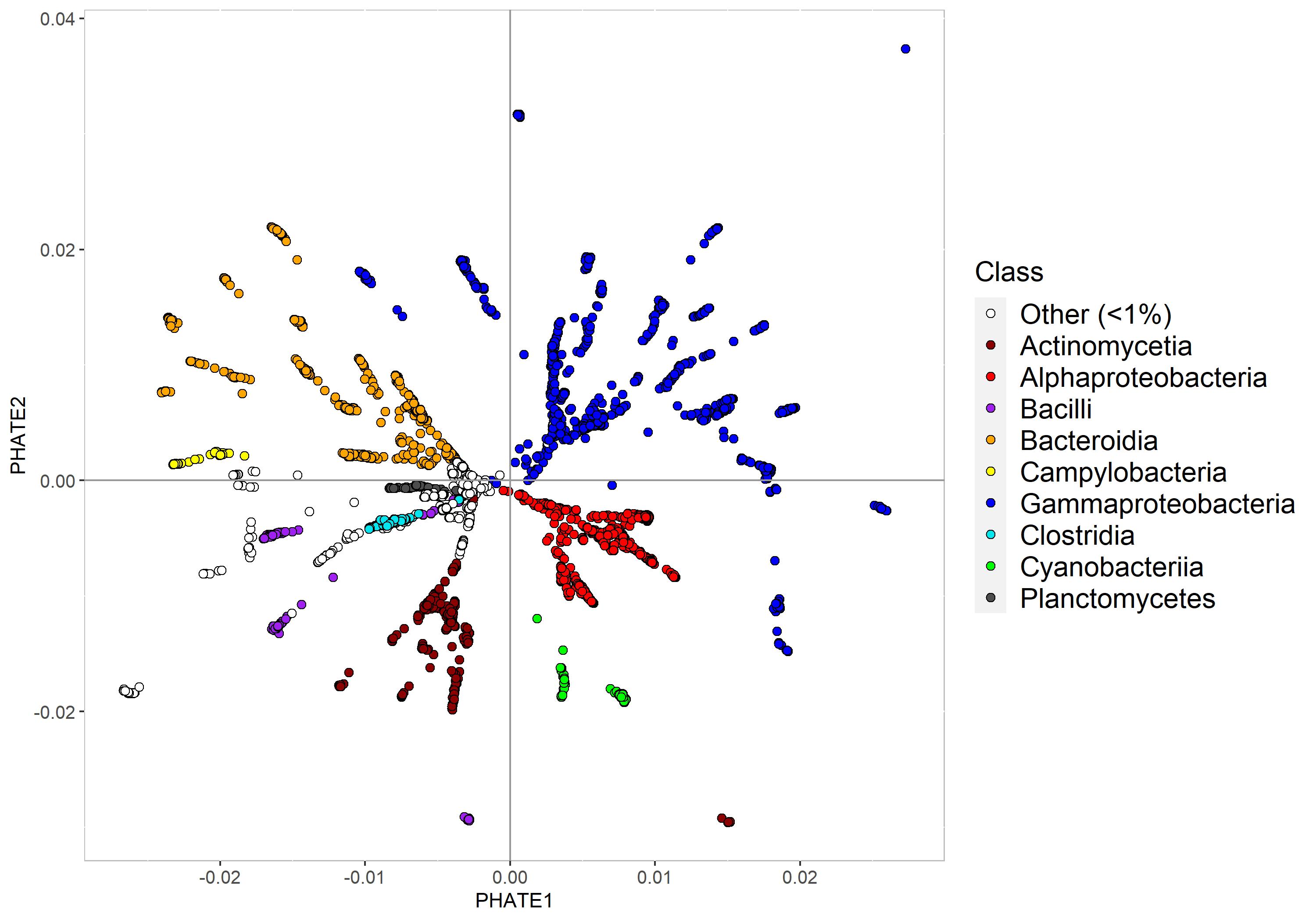}
    \caption{Two-dimensional embedding of diffusion variables created using the PHATE algorithm \cite{moon2019visualizing}. Each point represents an individual genome that is coloured by taxonomic class.}
    \label{phate_plot}
\end{figure}

The examples from above demonstrate that the diffusion map finds reasonable metabolic strategies of the bacterial community. Jointly these strategies define the metabolic niche space of the community and assign the taxa to specific coordinates in a multidimensional space. Using the PHATE procedure \cite{moon2019visualizing}, we combined the diffusion variables to embed the data geometry in a low-dimensional visualization of the metabolic niche space (Figure \ref{phate_plot}). Although it is important to interpret low-dimensional embeddings cautiously, as in previous work, the strategy space of the Baltic Sea bacterial community appears as a tree-like structure comprising clusters of taxa featuring localized strategies and continuous branches  \cite{fahimipour2020mapping} (Figure \ref{phate_plot}). 
The geometric structure implies large areas of the metabolic niche space that are unoccupied, reflecting either strategies that have yet to be discovered or strategies that are not feasible in the respective ecosystem or even in general \cite{fahimipour2020mapping}. 

The structure of the metabolic niche space roughly aligns with the phylogeny of the taxa. Uncultured taxa and the streamlined genomes of Patescibacter \cite{tian2020small}, Pelagibacterales \cite{grote2012streamlining} and Rickettsiales \cite{darby2007intracellular} form the core from which the branched structure emerges. The position of the streamlined genomes seems reasonable as these genomes retain mostly the key basic functions necessary for survival and reproduction that they share with many other organisms and they lack many of the more 'specialized' genes \cite{giovannoni2014implications}. The uncultured taxa group to the center either because they as well possess streamlined genomes or due to the lack of knowledge about their genes and respective functions. Therefore, their position in the metabolic niche space might change with further knowledge gained. 

Taxa that are characterized by localized variables such as the Cyanobacteria are more distinct within the structure. These distinct clusters could indicate that complex, costly adaptions are necessary to follow the respective metabolic strategy and the machinery cannot easily be acquired for example through horizontal gene transfer \cite{blankenship2010early}. Localized strategies could also reflect that an intermediate strategy is not feasible, for instance due to certain trade-offs resulting from adopting the respective strategy \cite{ferenci2016trade}.

Bacterial taxa associated with human diseases, like relatives of \textit{Klebsiella}, \textit{Mycobacterium}, \textit{Staphylococcus} and \textit{Fusobacterium}, group furthest away from the main structure and appear as clusters of dots in the periphery (Figure \ref{phate_plot}). The reason for these taxa to group away is probably the bias in global databases towards human pathogens and their functional capabilities (see above).

\subsection{Metabolic strategies over time. }
Above, we saw that the new variables identified by diffusion mapping represent interpretable metabolic strategies of the sampled community. Taking the relative abundances of the taxa that map to different genomes into account, we calculated the abundance-weighted means of each diffusion variable side for each sampling time-point (see \textsl{Methods}). This enables us to observe how the occupation of metabolic niches or strategies change over time in the Baltic Sea bacterial community. 

\begin{figure} [H]
   \centering
    \includegraphics[width=\textwidth]{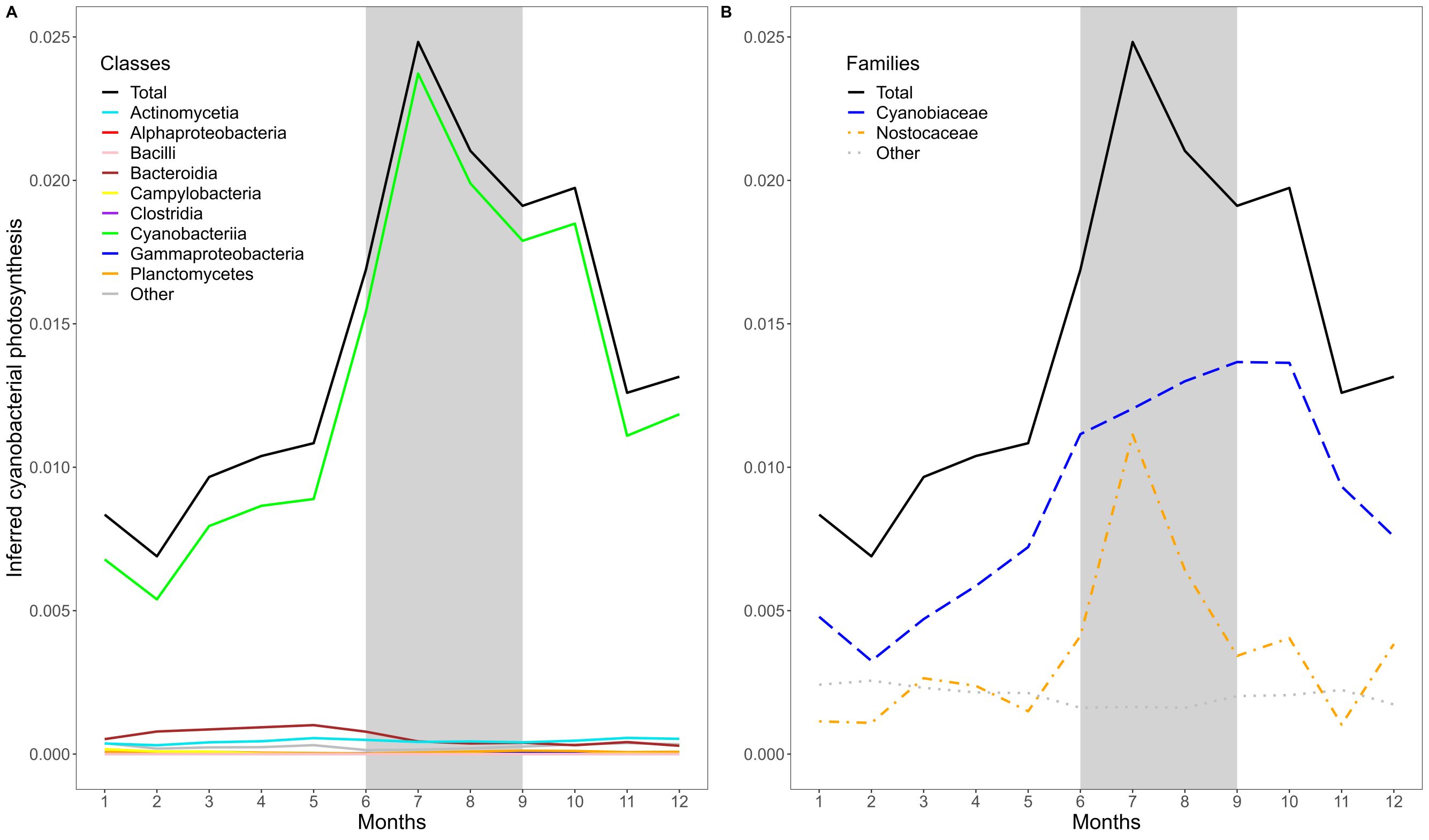}
    \caption{Abundance-weighted mean values of inferred cyanobacterial photosynthesis over the yearly cycle. Summer months are indicated by a gray background. Taxonomic class (A) and cyanobacterial taxonomic families (B) are color-coded.}.
    \label{PS_overtime}
\end{figure}

For example, the variable 4 negative, identified as cyanobacterial photosynthesis, reaches its highest abundance-weighted niche values in summer (Figure \ref{PS_overtime}). Cyanobacteria are known to cause massive summer blooms in the Baltic Sea \cite{lindh2018sensitivity}, supporting our interpretation that this strategy represents cyanobacterial photosynthesis. Separating the different cyanobacterial families reveals an early summer peak caused by the filamentous Nostocaceae and a plateau high of the niche values for the unicellular Cyanobiaceae increasing until the beginning of autumn (Figure \ref{PS_overtime}). Utilization of nutrients from filamentous Cyanobacteria might fuel the metabolism of opportunistic picocyanobacteria \cite{bertos2016unscrambling}.

\begin{figure} [H]
   \centering
    \includegraphics[width=\textwidth]{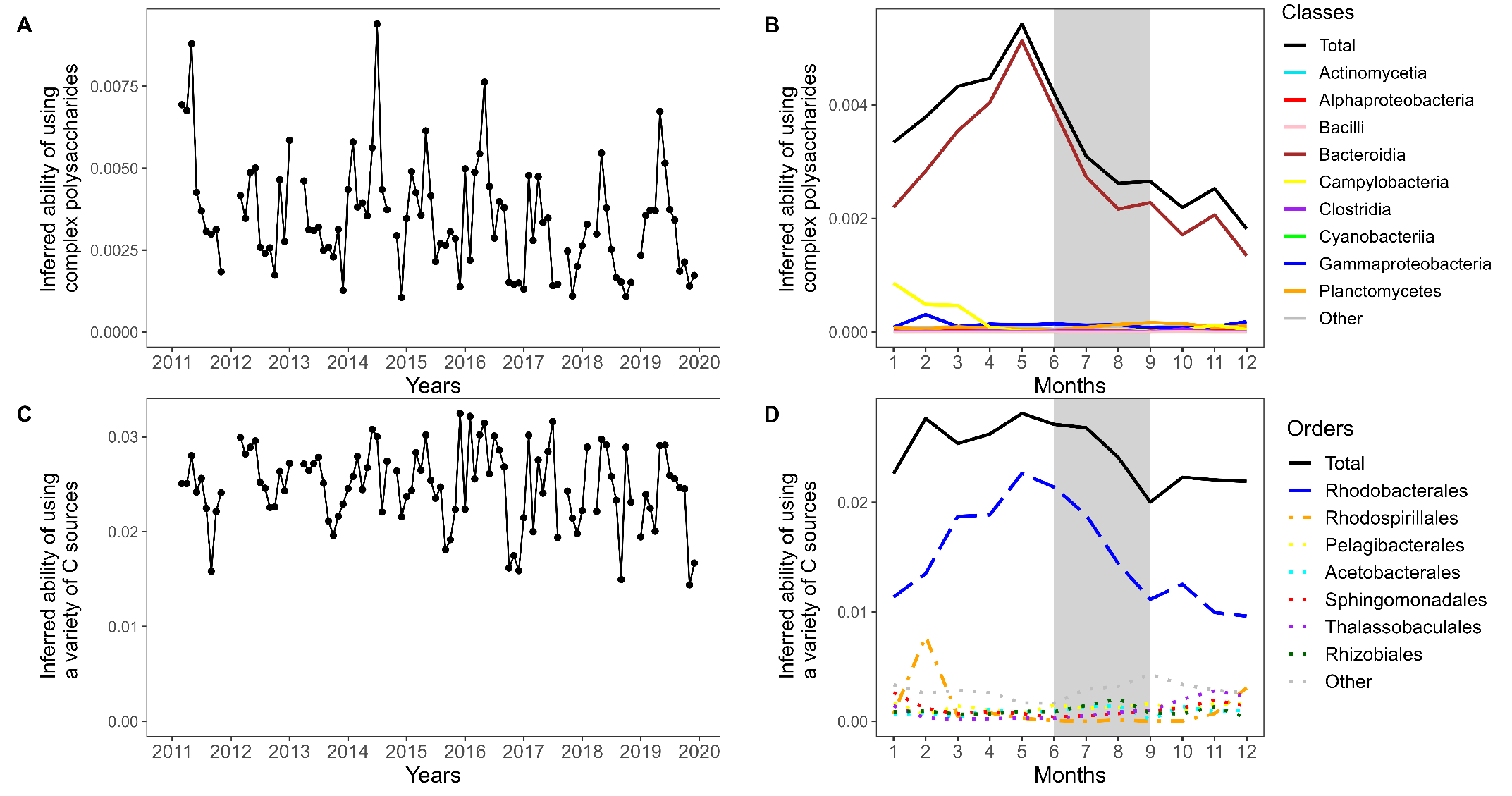}
    \caption{Abundance-weighted mean values of inferred ability of using complex polysaccharides (A, B) and inferred ability of using a variety of carbon sources(C,D) over the whole time period and over the yearly cycle. Summer months are indicated by a gray background. Taxonomic orders (B) and taxonomic classes (D) are color-coded.}.
    \label{EV3_14_over_time}
\end{figure}

Another strategy that is affected by seasonality is the catabolism of complex polysaccharides, identified in the variable 14 negative. Dominated by members of the Flavobacteriaceae, this strategy reaches highest abundance-weighted mean values in May (Figure~\ref{EV3_14_over_time}~B), following the peak of the phytoplankton spring bloom \cite{bunse2019high}. Large amounts of photosynthetic products, mainly polysaccharides, are exuded by phytoplankton \cite{muhlenbruch2018mini} and Flavobacteria are adapted to use these high-molecular-weight molecules \cite{buchan2014master}. In spring 2011 coinciding with the highest phytoplankton biomass \cite{bunse2019high}, this strategy reaches highest values, compared to all other sampling years (Figure~\ref{EV3_14_over_time}~A). 

Variable positive 3, describing the ability of utilizing a variety of carbon sources, also reaches highest values in May, but does not show a pronounced peak and instead decreases more slowly, reaching a minimum mean value in September (Figure~\ref{EV3_14_over_time}~D). This strategy is dominated by the marine Rhodobacteraceae, that are crucial in processing low-molecular-weight phytoplankton-derived metabolites and characterized by their high trophic versatility \cite{ferrer2021resource, newton2010genome}. In winter 2015/16 and 2016/17 elevated values of this variable positive 3 are driven by Rhodospirillales, especially of the genus \emph{Thalassospira} (Figure~\ref{EV3_14_over_time}~C). The latter is known for its ability to degrade polycylic aromatic hydrocarbons (PAHs) \cite{tsubouchi2014thalassospira} and their appearance in the upper water column could be related to Major Baltic Inflow events \cite{bergen2018impact} and subsequent winter mixing. 

In contrast to the strategies we discussed above that were positively impacted by the seasonal phytoplankton spring bloom, variable negative 38 reaches its minimum in May, right after the phytoplankton bloom (Supplementary Figure~11). Describing the metabolic ability to oxidize methyl groups and C1 compounds, such as methanol and formaldehyde, this strategy is dominated by Alphaproteobacteria, especially \emph{Pelagibacter} in winter and Planctomycetes in autumn in the Baltic Sea bacterial community. Marine dissolved organic carbon is a source for diverse C1 and methylated compounds, methanol constitutes a major fraction of oxygenated volatile organic chemicals and formaldehyde is omnipresent in seawater \cite{sun2011one}. The ability of using these compounds enables energy production from relatively abundant substrates in the water, but this ability is outcompeted when concentrations of phytoplankton-derived substrates increase in spring. Variable 43 positive describing non-spore forming sulfate reducers also reaches its minimum mean value in May (Supplementary Figure~12). Baltic Sea sulfate reducers of the phylum Desulfobacterota, inhabiting mostly sediments and oxygen-depleted waters \cite{vetterli2015seasonal, korneeva2015sulfate}, drive the peak of this strategy in February, probably appearing in the upper water column due to strong winter mixing. 
\newline 

\begin{figure} [H]
   \centering
    \includegraphics[width=\textwidth]{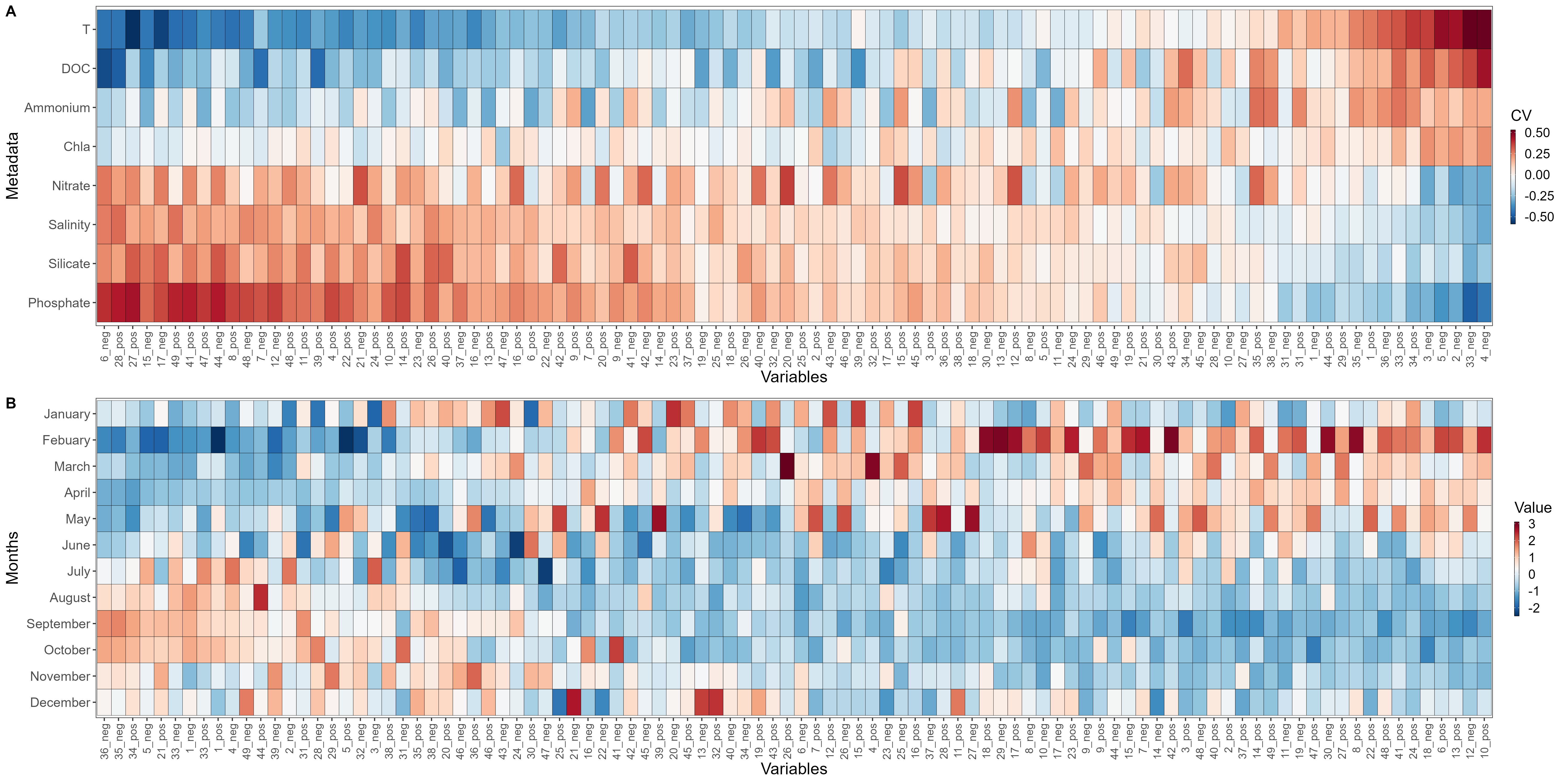}
    \caption{Heatmap of Spearman correlation coefficients (CV) between the first 49 variables, i.e. strategy time-series and the environmental variables (metadata) (A) and heatmap of abundance-weighted variable mean values of the first 49 variables for each month over the whole sampling period, standardized to mean = 0 and standard deviation = 1 for each variable side (B). T: temperature, DOC: dissolved organic carbon concentrations, Chla: chlorophyll a concentrations.
    }
    \label{heatmap}
\end{figure}

Comparing the strategy time-series to the environmental data obtained at LMO and the abundance-weighted mean values over the months, we can see a strong signal of seasonality (Figure~\ref{heatmap}): The variables on the left side of Figure \ref{heatmap}~A are correlated to rather high nutrient concentrations and lower temperatures, hence winter conditions, whereas the variables on the right side of the heatmap are correlated to higher temperatures and chlorophyll a concentrations, hence summer conditions. Figure \ref{heatmap}~B complements this picture: On the left side we find higher variable values in summer and autumn, while the variables on the right hand side show higher values in winter and spring.

\subsection{Functional diversity. }
\begin{figure} [H]
   \centering
    \includegraphics[width=\textwidth]{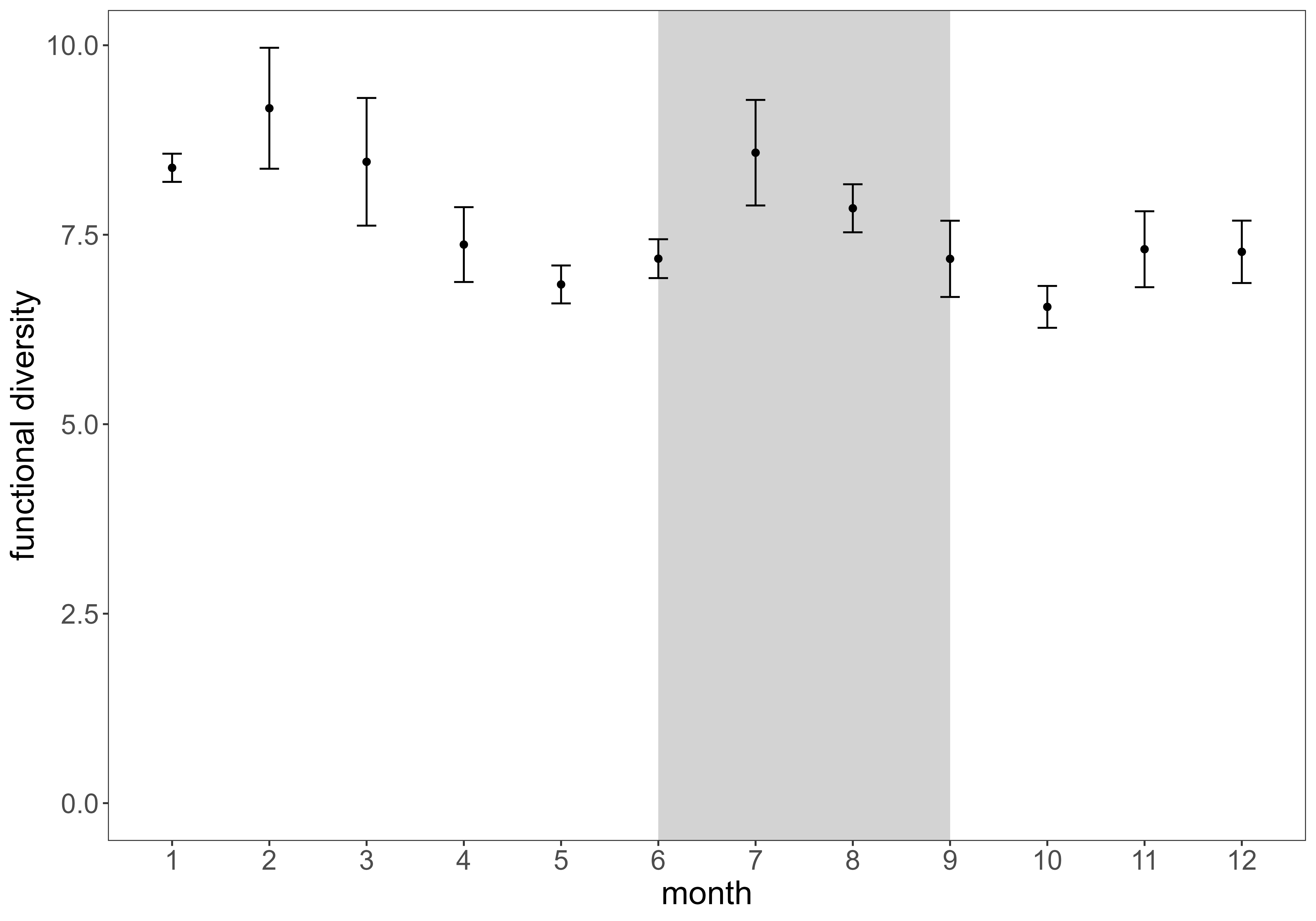}
    \caption{Mean and standard error of functional diversity estimation calculated as Rao index \cite{ryabov2021trait} for the sampled Baltic Sea bacterial community over the yearly cycle summarizing the years 2011-2019. Gray background indicates summer months.}.
    \label{FuncDiv_months}
\end{figure}

From the variables obtained via diffusion mapping we derived the diffusion distances between all pairs of species to quantify the functional diversity calculated as Rao index \cite{ryabov2021trait} (see Methods). The yearly cycle summarized over all sampling years shows highest functional diversity mean values in February and July, whereas it reaches lowest values in May and October (Fig. \ref{FuncDiv_months}). This pattern is likely explained by the absence of the thermocline in winter, that causes deeper mixing of the water layers \cite{reissmann2009vertical}, resulting in bacterial communities from the former mesopelagic to be found also in surface waters \cite{herlemann2016phylogenetic, lindh2015disentangling}. Members of these communities can possess very different strategies due to their adaptation to deeper water layer or sediment conditions \cite{lindh2015disentangling} and their taxonomic and functional diversity is in general higher than in surface communities \cite{sunagawa2015structure}, hence they increase the functional diversity when they are observed in the surface layers in winter. Functional diversity also benefits from increased diversity of nutrients available in winter, leading to an increased resource heterogeneity \cite{rieck2015particle, raes2022seasonal}. Lower values of functional diversity in May, following the phytoplankton bloom, could be explained by the strong dominance of strategies that relate to the utilization of phytoplankton-derived substrates during bloom phases as these compounds increase drastically in relative abundance \cite{teeling2012substrate, buchan2014master, pernthaler2017competition}. 

\section{Conclusion}
In this paper we showed that relative abundance of prokaryotic populations obtained from amplicon sequencing data from monitoring datasets can be translated into potentially occupied ecological niches over time. The diffusion map of the bacterial capabilities detected a wide spectrum of interpretable metabolic strategies of the Baltic Sea bacterial community: From localized strategies such as cyanobacterial photosynthesis to continua of strategies such as degree of association with marine hosts and degree of trophic versatility, revealing both, strategies that align with phylogeny, strategies that differentiate closely-related taxa, and similar strategies among distantly-related taxa \cite{fahimipour2020mapping}. The latter may reflect metabolic niche convergence \cite{pianka2017toward} or horizontal gene transfer \cite[e.g.][]{gogarten2005horizontal}. 

Systematizing the genomic information via our diffusion map approach provided the possibility to express the changes in bacterial species abundances as quantitative changes in potential occupation of metabolic niches over time. These here called abundance-weighted strategy values showed a variety of patterns in time: Strategies displaying seasonal dynamics, e.g. increasing trends in summer or an increase following the phytoplankton bloom or higher values as a consequence of winter mixing as well as strategies showing interannual changes in time patterns. Impacts of single events, for example a very pronounced spring bloom or a Major Baltic Inflow event, were also reflected in the strategy time-series. Some functional strategies are clearly dominated by one bacterial group, whereas others are divided between bacterial groups often depending on season. Overall, seasonality seems to be a strong driver not only for phylogenetic bacterial composition \cite{lindh2015disentangling, herlemann2016phylogenetic} but also for bacterial metabolic niche occupation and functional diversity of bacterial communities in the Baltic Sea. Seasonality is probably such an important driver also for metabolic strategies due to the interplay of seasonal changes in substrate availabilities together with changes in abiotic parameters that influence metabolic activities \cite{ward2017annual, seidel2017composition, herlemann2016phylogenetic}.

The first new variable identified by the diffusion map separates the metabolic strategies of pathogenic members of the Enterobacteriaceae from all other taxa. The reason for this variable showing up first is that the diffusion map probably detects the bias that bacterial pathogenic taxa and genes involved in pathogenesis are overrepresented in global databases \cite[e.g.][]{blackwell2021exploring}. This makes clear that a major limitation of this approach is our understanding of genes and their functions, implying also that the metabolic niche space might change due to further knowledge gained. For example, we might observe more branches representing currently undetected taxa or strategies and taxa currently grouped towards the center might shift further outwards and distances between taxa might change with increasing information. On the other hand this result also highlights the power of the diffusion map approach to objectively detect such biases in the data.  

It should be noted that genes only encode the theoretical capabilities of a species \cite{fahimipour2020mapping}, conceptually corresponding to the fundamental niche concept \cite{hutchinson1957cold}. It was however shown that functional genes can be used to predict the position of species along major niche gradients, outperforming the predictions based on phylogenetic information \cite{alneberg2020ecosystem}. Gowda et al. \cite{gowda2022genomic} also showed for the process of denitrification that it is possible to predict community metabolic dynamics from the presence and absence of genes in metagenomes. In comparison to their studies, our approach of assigning ASVs to species and obtaining their complete genomes from a database relies heavily on the quality of the databases and genome assemblages available. To overcome these shortcomings, the analysis of metagenomes is a desirable tool to improve our analysis by accounting also for within-species genome variability, the accessory genome and prophages that can play a role in shaping the strategies and traits of the respective organisms \cite{garcia2019microdiversity, zhu2015inter, forcone2021prophage}. Ideally, we would have all the genomes available from exactly the taxa from the habitat sampled, as we do not, we rely on a simple mapping scheme. In the future, deep shotgun metagenomic sequencing and long read technologies might be important for collecting the sorts of data that would make our method even more powerful.

We conclude that the diffusion map approach presented here enables us to coarse grain complex bacterial communities in terms of reasonable metabolic strategies and provides a quantitative framework to organize genomic information into potentially occupied metabolic niches over time. Thereby, this approach enables us to understand dynamics of community composition on different scales in terms of their impact on potentially occupied metabolic niches and to link genomic data to metabolic strategies enhancing our understanding of the relationships between community compositions and ecosystem functions.

\section{Materials and Methods}
\subsection{Sampling data. }
Seawater samples were obtained from the Linnaeus Microbial Observatory (LMO) (N 56\textdegree55.854', E 17\textdegree3.6420') situated in the Western Baltic Proper approximately weekly during 2011-2013 and monthly during 2014-2019. Using 3 or 5~l Ruttner water samplers, water was sampled from 2~m depth at at $\sim$ 9~a.m. during each sampling occasion.
Seawater was processed in the laboratory (Linnaeus University) where environmental parameters (temperature, salinity, chlorophyll a, DOC, nitrate and nitrite (together named as nitrate), phosphate, silicate and ammonium were analyzed as previously described \cite{lindh2015disentangling, bunse2019high}, Fridolfsson et al. in prep.). 
For microbial community composition, we filtered the seawater directly onto 0.22~{\textmu}m SterivexTM cartridge filters (Millipore), or prefiltered onto 3~{\textmu}m polycarbonate filters and subsequently on 0.22~{\textmu}m SterivexTM cartridges (named 3-0.2~{\textmu}m size fraction) using a persistaltic pump. 
We stored the filters in TE buffer at -80°C until DNA extraction using a phenol-chloroform method described by Boström et al. \cite{bostrom2004optimization} and modified after Bunse et al. \cite{bunse2016spatio}. We amplified the V3V4 region of the 16S rRNA gene using PCRs with the primer pair 341f-805r \cite{herlemann2011transitions, hugerth2014degeprime}. We analyzed the DNA concentrations with a Nanodrop or Qubit 2.0 Fluorometer (Life Technologies) and ran gel electrophoresis to confirm the amplicon specificity. 
Sample batches for sequencing were successively sent to the Science for Life Laboratory, Sweden on the Illumina MiSeq platform, resulting in 2 × 300 bp paired-end reads. For bioinformatic processing, we used the nf-core Ampliseq pipeline \cite{ewels2020nf, straub2020interpretations} with the following software versions:  nf-core/ampliseq = v1.2.0dev; Nextflow = v20.10.0; FastQC = v0.11.8; MultiQC = v1.9; Cutadapt = v2.8; QIIME2 = v2019.10.0. We used DADA2 \cite{callahan2016dada2} implemented in QIIME2 \cite{bolyen2019reproducible} and trimmed the sequences at forward 259 bp and reverse 199 bp before denoising. 
Of all LMO samples, we used all filter fractions for the niche space analysis but only the non-prefiltered 0.22~µm fraction for abundance estimates (see method description below). The LMO 16S rRNA raw data used in this study are deposited in the EMBL-EBI European Nucleotide Archive repository (https://www.ebi. ac.uk/ena) under accession numbers PRJEB52828, PRJEB52855, SRP048666, PRJEB52837, PRJEB40890, PRJEB52851, PRJEB52854, PRJEB52627, PRJEB52772, PRJEB52496, PRJEB52780, PRJEB52782, PRJEB52850.

\subsection{Obtaining genomes and genes from ASV data. }
We conducted a BLAST sequence similarity search to match the ASV sequences obtained from 16S rRNA analysis from the time-series to the Genome Taxonomy Database (GTDB, \url{https://data.gtdb.ecogenomic.org/}, release 95 database). 21,102 (so 44\%) of the 48,098 ASVs,  could be matched well (similarity greater than 95\%) to a genome from the GTDB. In terms of abundances, a mean of 82\% (column sums) were good matches. From the well matched species we obtained the complete genomes from GTDB and NCBI (Refseq and GenBank). These 4,265 complete genomes were annotated using Prokka \cite{seemann2014prokka}.
 
\subsection{Diffusion mapping the strategy space. }
We diffusion mapped the genomes according to their similarity in known gene composition (absence-presence of genes) using the algorithm described in \cite{barter2019manifold}, applying the inverse of the hamming distance as a similarity measure. This method briefly consists of five steps. The starting point is the data matrix A with the dimensions $M$ x $N$, where $M$=4,265 is the number of genomes and $N$=15,361 is the number of annotated genes. First, the data is standardized so that each column has a mean of zero and a standard deviation of 1. Each row of our standardized data matrix can be interpreted as a vector of coordinates of the respective genome in a $N$-dimensional space. Second, we define similarities of two genomes as the inverse of the hamming distance between these vectors. The computed similarities form a $M$x$M$ matrix, in which we set the diagonal elements to zero. We interpret the results as the weight matrix of the network with high values indicating close similarities between the respective genomes in terms of known gene composition. Third, we threshold the weight matrix, keeping only the top-25 highest similarities for each genome. From this weighted adjacency matrix we compute the corresponding row-normalized Laplacian matrix (step 4). In a final step, the eigenvectors and corresponding eigenvalues of this Laplacian are computed, defining new variables describing important variation in the dataset. The diffusion map identifies new variables spanning the strategy space of the analyzed bacteria and archaea and orders these variables according to their importance. Every genome is assigned a coordinate entry for each variable, so that we can order the corresponding taxa according to their entry, from most negative to most positive entries. Strategies of taxa that are assigned extreme entries along a variable can help interpret the meaning of each variable. 

\subsection{Translating ASV time-series into strategy time-series. }
For analysis, we separated the negative and positive values for each diffusion variable. To convert the species time-series into a strategy time-series, we calculated the weighted means of each diffusion variable for each sampling time-point, multiplying the eigenvector entries of each taxa by the normalized abundance of this taxa at this time-point. 

\subsection{Identifying over-represented genes. }
 To identify the genes that were over-re\-pre\-sen\-ted in the genomes of the taxa that themselves were assigned extreme entries along diffusion map variables, we used a permutational variant of the gene set enrichment analysis, GSEA \cite{subramanian2005gene}. Genomes were ranked by the orderings specified by each diffusion variable. Gene set enrichment analysis was performed using the fgsea library in R \cite{R,fgsea} with a Benjamini-Hochberg-adjusted \cite{benjamini1995controlling} $P$ value \textless 0.01 used as a threshold for retaining genes corresponding to taxa that scored extreme values in the respective diffusion variable. 
 
\subsection{Estimating functional diversity. }
Functional diversity was estimated using the procedure described by Ryabov et al. \cite{ryabov2021trait}. Briefly, the method uses the euclidean distances between species in the strategy space, rescaling the eigenvector elements according to their respective eigenvalue to quantify pairwise functional dissimilarities between all species. These newly defined diffusion distances can then be used to obtain the functional diversity of each sample, calculated as the Rao index. 

\subsection{Data processing. }
Data was processed with R~4.1.2 \cite{R} and Julia~1.7.1 \cite{Julia-2017}. The two dimensional mapping of the diffusion variables was made using the package phateR \cite{moon2019visualizing}. Also the R packages ggplot2 \cite{ggplot2}, tidyverse \cite{tidyverse}, plyr \cite{plyr}, ggbreak \cite{ggbreak}, ggpubr \cite{ggpubr}, seriation \cite{seriation} and reshape \cite{reshape} were used.

\section*{Supplemental Material file list}

\textbf{Supplementary Tables}
\begin{enumerate}[label = S\arabic*]
    \item Genomes that map to the 100 most abundant ASVs.
    \item Species that map to the 100 ASVs scoring most negative values in Variable~1.
    \item Top 100 over-represented annotated genes in the genomes of the taxa that receive the most negative entries on variable 1.
    \item Top 100 over-represented annotated genes in the genomes of the taxa that receive the most positive entries on variable 2.
    \item Top 100 over-represented annotated genes in the genomes of the taxa that receive the most positive entries on variable 3.
    \item Top 100 over-represented annotated genes in the genomes of the taxa that receive the most negative entries on variable 4. 
    \item Top 100 over-represented annotated genes in the genomes of the taxa that receive the most negative entries on variable 14.
    \item Top 100 over-represented annotated genes in the genomes of the taxa that receive the most negative entries on variable 38. 
    \item Top 100 over-represented annotated genes in the genomes of the taxa that receive the most positive entries on variable 43. 
\end{enumerate}

\textbf{Supplementary Figures}
\begin{enumerate}[label = S\arabic*]
    \item Relative mean abundances of classes that map to the genomes obtained from amplicon sequencing data over the whole sampling period.
    \item The ordering of taxa defined by variable 3 entries.
    \item The ordering of taxa defined by variable 4 entries. 
    \item The ordering of taxa defined by variable 14 entries.
    \item The ordering of taxa defined by variable 27 entries. 
    \item The ordering of taxa defined by variable 33 entries. 
    \item The ordering of taxa defined by variable 38 entries. 
    \item The ordering of taxa defined by variable 43 entries. 
    \item Abundance-weighted mean values of inferred ability of utilizing a variety of carbon sources over the yearly cycle.
    \item Abundance-weighted mean values of inferred ability of degrading complex polysaccharides over the yearly cycle.
    \item Abundance-weighted mean values of inferred ability of oxidizing methyl groups and C1 compounds over the yearly cycle.
    \item Abundance-weighted mean values of trait dominated by non-spore forming sulfate reducers over the yearly cycle.
\end{enumerate}

\section*{Acknowledgments}
We thank the many people involved in LMO sampling and lab work over the years, especially Sabina Arnautovic, Anders Månsson, Emil Fridolfsson, Benjamin Pontiller and Kristofer Bergström. Furthermore, we acknowledge the help and assistance from Northern Offshore Services (NOS), M/V Provider crew, E.ON and RWE on the samplings. We acknowledge Daniel Lundin for bioinformatic support. The LMO research was supported by the Swedish Research Council FORMAS Strong Research environment EcoChange (Ecosystem dynamics in the Baltic Sea in a changing climate) to JP. CB, JM and TG were supported by the Helmholtz Institute for Functional Marine Biodiversity (HIFMB), a collaboration between the Alfred-Wegener-Institute, Helmholtz-Center for Polar and Marine Research, and the Carl-von-Ossietzky University Oldenburg, initially funded by the Ministry of Science and Culture of Lower Saxony (HIFMB Project) and the Volkswagen Foundation through the “Niedersächsisches Vorab” grant program (grant number ZN3285).We acknowledge support by the Open Access Publication Funds of Alfred-Wegener-Institut Helmholtz-
Zentrum für Polar- und Meeresforschung.

\bibliographystyle{unsrtnat}

\begin{thebibliography}{94}
\providecommand{\natexlab}[1]{#1}
\providecommand{\url}[1]{\texttt{#1}}
\expandafter\ifx\csname urlstyle\endcsname\relax
  \providecommand{\doi}[1]{doi: #1}\else
  \providecommand{\doi}{doi: \begingroup \urlstyle{rm}\Url}\fi

\bibitem[Hutchinson(1961)]{hutchinson1961paradox}
G~Evelyn Hutchinson.
\newblock The paradox of the plankton.
\newblock \emph{The American Naturalist}, 95\penalty0 (882):\penalty0 137--145,
  1961.

\bibitem[Locey and Lennon(2016)]{locey2016scaling}
Kenneth~J Locey and Jay~T Lennon.
\newblock Scaling laws predict global microbial diversity.
\newblock \emph{Proceedings of the National Academy of Sciences}, 113\penalty0
  (21):\penalty0 5970--5975, 2016.

\bibitem[Louca et~al.(2019)Louca, Mazel, Doebeli, and Parfrey]{louca2019census}
Stilianos Louca, Florent Mazel, Michael Doebeli, and Laura~Wegener Parfrey.
\newblock A census-based estimate of earth's bacterial and archaeal diversity.
\newblock \emph{PLoS biology}, 17\penalty0 (2):\penalty0 e3000106, 2019.

\bibitem[Sunagawa et~al.(2015)Sunagawa, Coelho, Chaffron, Kultima, Labadie,
  Salazar, Djahanschiri, Zeller, Mende, Alberti, et~al.]{sunagawa2015structure}
Shinichi Sunagawa, Luis~Pedro Coelho, Samuel Chaffron, Jens~Roat Kultima,
  Karine Labadie, Guillem Salazar, Bardya Djahanschiri, Georg Zeller, Daniel~R
  Mende, Adriana Alberti, et~al.
\newblock Structure and function of the global ocean microbiome.
\newblock \emph{Science}, 348\penalty0 (6237):\penalty0 1261359, 2015.

\bibitem[Zhang et~al.(2019)Zhang, Ding, Li, Tam, Bougouffa, Wang, Pei, Chiang,
  Leung, Lu, et~al.]{zhang2019marine}
Weipeng Zhang, Wei Ding, Yong-Xin Li, Chunkit Tam, Salim Bougouffa, Ruojun
  Wang, Bite Pei, Hoyin Chiang, Pokman Leung, Yanhong Lu, et~al.
\newblock Marine biofilms constitute a bank of hidden microbial diversity and
  functional potential.
\newblock \emph{Nature communications}, 10\penalty0 (1):\penalty0 1--10, 2019.

\bibitem[Falkowski et~al.(2008)Falkowski, Fenchel, and
  Delong]{falkowski2008microbial}
Paul~G Falkowski, Tom Fenchel, and Edward~F Delong.
\newblock The microbial engines that drive earth's biogeochemical cycles.
\newblock \emph{science}, 320\penalty0 (5879):\penalty0 1034--1039, 2008.

\bibitem[Hadfield(2011)]{hadfield2011biofilms}
Michael~G Hadfield.
\newblock Biofilms and marine invertebrate larvae: what bacteria produce that
  larvae use to choose settlement sites.
\newblock \emph{Annual review of marine science}, 3:\penalty0 453--470, 2011.

\bibitem[Bourne et~al.(2016)Bourne, Morrow, and Webster]{bourne2016insights}
David~G Bourne, Kathleen~M Morrow, and Nicole~S Webster.
\newblock Insights into the coral microbiome: underpinning the health and
  resilience of reef ecosystems.
\newblock \emph{Annual Review of Microbiology}, 70:\penalty0 317--340, 2016.

\bibitem[Bunse and Pinhassi(2017)]{bunse2017marine}
Carina Bunse and Jarone Pinhassi.
\newblock Marine bacterioplankton seasonal succession dynamics.
\newblock \emph{Trends in microbiology}, 25\penalty0 (6):\penalty0 494--505,
  2017.

\bibitem[Fuhrman et~al.(2015)Fuhrman, Cram, and Needham]{fuhrman2015marine}
Jed~A Fuhrman, Jacob~A Cram, and David~M Needham.
\newblock Marine microbial community dynamics and their ecological
  interpretation.
\newblock \emph{Nature Reviews Microbiology}, 13\penalty0 (3):\penalty0
  133--146, 2015.

\bibitem[Buttigieg et~al.(2018)Buttigieg, Fadeev, Bienhold, Hehemann, Offre,
  and Boetius]{buttigieg2018marine}
Pier~Luigi Buttigieg, Eduard Fadeev, Christina Bienhold, Laura Hehemann, Pierre
  Offre, and Antje Boetius.
\newblock Marine microbes in 4d—using time series observation to assess the
  dynamics of the ocean microbiome and its links to ocean health.
\newblock \emph{Current opinion in microbiology}, 43:\penalty0 169--185, 2018.

\bibitem[Auladell et~al.(2021)Auladell, Barber{\'a}n, Logares, Garc{\'e}s,
  Gasol, and Ferrera]{auladell2021seasonal}
Adri{\`a} Auladell, Albert Barber{\'a}n, Ramiro Logares, Esther Garc{\'e}s,
  Josep~M Gasol, and Isabel Ferrera.
\newblock Seasonal niche differentiation among closely related marine bacteria.
\newblock \emph{The ISME Journal}, pages 1--12, 2021.

\bibitem[Needham and Fuhrman(2016)]{needham2016pronounced}
David~M Needham and Jed~A Fuhrman.
\newblock Pronounced daily succession of phytoplankton, archaea and bacteria
  following a spring bloom.
\newblock \emph{Nature microbiology}, 1\penalty0 (4):\penalty0 1--7, 2016.

\bibitem[Bellman(1957)]{bellman1957dynamic}
Richard Bellman.
\newblock Dynamic programming.
\newblock \emph{Princeton, USA: Princeton University Press}, 1\penalty0
  (2):\penalty0 3, 1957.

\bibitem[Moon et~al.(2019)Moon, van Dijk, Wang, Gigante, Burkhardt, Chen, Yim,
  van~den Elzen, Hirn, Coifman, et~al.]{moon2019visualizing}
Kevin~R Moon, David van Dijk, Zheng Wang, Scott Gigante, Daniel~B Burkhardt,
  William~S Chen, Kristina Yim, Antonia van~den Elzen, Matthew~J Hirn, Ronald~R
  Coifman, et~al.
\newblock Visualizing structure and transitions in high-dimensional biological
  data.
\newblock \emph{Nature biotechnology}, 37\penalty0 (12):\penalty0 1482--1492,
  2019.

\bibitem[Barter and Gross(2019)]{barter2019manifold}
Edmund Barter and Thilo Gross.
\newblock Manifold cities: social variables of urban areas in the uk.
\newblock \emph{Proceedings of the Royal Society A}, 475\penalty0
  (2221):\penalty0 20180615, 2019.

\bibitem[Ghafourian et~al.(2020)Ghafourian, Georgiou, Barter, and
  Gross]{ghafourian2020wireless}
Amin Ghafourian, Orestis Georgiou, Edmund Barter, and Thilo Gross.
\newblock Wireless localization with diffusion maps.
\newblock \emph{Scientific Reports}, 10\penalty0 (1):\penalty0 1--10, 2020.

\bibitem[Ma and Fu(2012)]{ma2012manifold}
Yunqian Ma and Yun Fu.
\newblock \emph{Manifold learning theory and applications}, volume 434.
\newblock CRC press Boca Raton, 2012.

\bibitem[Jolliffe and Cadima(2016)]{jolliffe2016principal}
Ian~T Jolliffe and Jorge Cadima.
\newblock Principal component analysis: a review and recent developments.
\newblock \emph{Philosophical Transactions of the Royal Society A:
  Mathematical, Physical and Engineering Sciences}, 374\penalty0
  (2065):\penalty0 20150202, 2016.

\bibitem[Coifman et~al.(2005)Coifman, Lafon, Lee, Maggioni, Nadler, Warner, and
  Zucker]{coifman2005geometric}
Ronald~R Coifman, Stephane Lafon, Ann~B Lee, Mauro Maggioni, Boaz Nadler,
  Frederick Warner, and Steven~W Zucker.
\newblock Geometric diffusions as a tool for harmonic analysis and structure
  definition of data: Diffusion maps.
\newblock \emph{Proceedings of the national academy of sciences}, 102\penalty0
  (21):\penalty0 7426--7431, 2005.

\bibitem[Coifman and Lafon(2006)]{coifman2006diffusion}
Ronald~R Coifman and St{\'e}phane Lafon.
\newblock Diffusion maps.
\newblock \emph{Applied and computational harmonic analysis}, 21\penalty0
  (1):\penalty0 5--30, 2006.

\bibitem[Fahimipour and Gross(2020)]{fahimipour2020mapping}
Ashkaan~K Fahimipour and Thilo Gross.
\newblock Mapping the bacterial metabolic niche space.
\newblock \emph{Nature communications}, 11\penalty0 (1):\penalty0 4887, 2020.

\bibitem[Ryabov et~al.(2021)Ryabov, Blasius, Hillebrand, Olenina, and
  Gross]{ryabov2021trait}
Alexey Ryabov, Bernd Blasius, Helmut Hillebrand, Irina Olenina, and Thilo
  Gross.
\newblock Trait reconstruction and estimation of functional diversity from
  ecolocical monitoring data.
\newblock \emph{arXiv preprint arXiv:2107.13792}, 2021.

\bibitem[Reiter et~al.(2021)Reiter, Montpetit, Runnebaum, Brown, and
  Montpetit]{reiter2021charting}
Taylor Reiter, Rachel Montpetit, Ron Runnebaum, C~Titus Brown, and Ben
  Montpetit.
\newblock Charting shifts in saccharomyces cerevisiae gene expression across
  asynchronous time trajectories with diffusion maps.
\newblock \emph{Mbio}, 12\penalty0 (5):\penalty0 e02345--21, 2021.

\bibitem[Hutchinson(1957)]{hutchinson1957cold}
G~Evelyn Hutchinson.
\newblock Cold spring harbor symposium on quantitative biology.
\newblock \emph{Concluding remarks}, 22:\penalty0 415--427, 1957.

\bibitem[Nyberg et~al.(2015)Nyberg, Gross, and Bassler]{nyberg2015mesoscopic}
Amy Nyberg, Thilo Gross, and Kevin~E Bassler.
\newblock Mesoscopic structures and the laplacian spectra of random geometric
  graphs.
\newblock \emph{Journal of Complex Networks}, 3\penalty0 (4):\penalty0
  543--551, 2015.

\bibitem[Anderson(1958)]{anderson1958absence}
Philip~W Anderson.
\newblock Absence of diffusion in certain random lattices.
\newblock \emph{Physical review}, 109\penalty0 (5):\penalty0 1492, 1958.

\bibitem[Pollack and Neilands(1970)]{pollack1970enterobactin}
J~Ri Pollack and JB~Neilands.
\newblock Enterobactin, an iron transport compound from salmonella typhimurium.
\newblock \emph{Biochemical and biophysical research communications},
  38\penalty0 (5):\penalty0 989--992, 1970.

\bibitem[O'brien and Gibson(1970)]{o1970structure}
IG~O'brien and F~Gibson.
\newblock The structure of enterochelin and related 2, 3-dihydroxy-n-benzoyne
  conjugates from eschericha coli.
\newblock \emph{Biochimica et Biophysica Acta (BBA)-General Subjects},
  215\penalty0 (2):\penalty0 393--402, 1970.

\bibitem[Aldridge and Hughes(2002)]{aldridge2002regulation}
Phillip Aldridge and Kelly~T Hughes.
\newblock Regulation of flagellar assembly.
\newblock \emph{Current opinion in microbiology}, 5\penalty0 (2):\penalty0
  160--165, 2002.

\bibitem[Domka et~al.(2006)Domka, Lee, and Wood]{domka2006ylih}
Joanna Domka, Jintae Lee, and Thomas~K Wood.
\newblock Ylih (bssr) and ycep (bsss) regulate escherichia coli k-12 biofilm
  formation by influencing cell signaling.
\newblock \emph{Applied and environmental microbiology}, 72\penalty0
  (4):\penalty0 2449--2459, 2006.

\bibitem[Blackwell et~al.(2021)Blackwell, Hunt, Malone, Lima, Horesh, Alako,
  Thomson, and Iqbal]{blackwell2021exploring}
Grace~A Blackwell, Martin Hunt, Kerri~M Malone, Leandro Lima, Gal Horesh,
  Blaise~TF Alako, Nicholas~R Thomson, and Zamin Iqbal.
\newblock Exploring bacterial diversity via a curated and searchable snapshot
  of archived dna sequences.
\newblock \emph{PLoS biology}, 19\penalty0 (11):\penalty0 e3001421, 2021.

\bibitem[Hallenbeck(2017)]{hallenbeck2017modern}
Patrick~C Hallenbeck.
\newblock \emph{Modern topics in the phototrophic prokaryotes: Metabolism,
  bioenergetics, and omics}.
\newblock Springer, 2017.

\bibitem[Raven(2011)]{raven2011cost}
John~A Raven.
\newblock The cost of photoinhibition.
\newblock \emph{Physiologia Plantarum}, 142\penalty0 (1):\penalty0 87--104,
  2011.

\bibitem[Luo and Moran(2015)]{luo2015divergent}
Haiwei Luo and Mary~Ann Moran.
\newblock How do divergent ecological strategies emerge among marine
  bacterioplankton lineages?
\newblock \emph{Trends in microbiology}, 23\penalty0 (9):\penalty0 577--584,
  2015.

\bibitem[Alves et~al.(2014)Alves, de~Souza, de~Mello~Varani, and
  de~Macedo~Lemos]{alves2014family}
Lucia Maria~Carrareto Alves, Jackson Ant{\^o}nio~Marcondes de~Souza, Alessandro
  de~Mello~Varani, and EG~de~Macedo~Lemos.
\newblock The family rhizobiaceae.
\newblock \emph{The Prokaryotes}, pages 419--437, 2014.

\bibitem[Zecher et~al.(2020)Zecher, Hayes, and Philipp]{zecher2020evidence}
Karsten Zecher, Kristiane~Rebecca Hayes, and Bodo Philipp.
\newblock Evidence of interdomain ammonium cross-feeding from methylamine-and
  glycine betaine-degrading rhodobacteraceae to diatoms as a widespread
  interaction in the marine phycosphere.
\newblock \emph{Frontiers in microbiology}, 11, 2020.

\bibitem[Tett et~al.(2021)Tett, Pasolli, Masetti, Ercolini, and
  Segata]{tett2021prevotella}
Adrian Tett, Edoardo Pasolli, Giulia Masetti, Danilo Ercolini, and Nicola
  Segata.
\newblock Prevotella diversity, niches and interactions with the human host.
\newblock \emph{Nature Reviews Microbiology}, 19\penalty0 (9):\penalty0
  585--599, 2021.

\bibitem[Buchan et~al.(2014)Buchan, LeCleir, Gulvik, and
  Gonz{\'a}lez]{buchan2014master}
Alison Buchan, Gary~R LeCleir, Christopher~A Gulvik, and Jos{\'e}~M
  Gonz{\'a}lez.
\newblock Master recyclers: features and functions of bacteria associated with
  phytoplankton blooms.
\newblock \emph{Nature Reviews Microbiology}, 12\penalty0 (10):\penalty0
  686--698, 2014.

\bibitem[Teeling et~al.(2012)Teeling, Fuchs, Becher, Klockow, Gardebrecht,
  Bennke, Kassabgy, Huang, Mann, Waldmann, et~al.]{teeling2012substrate}
Hanno Teeling, Bernhard~M Fuchs, D{\"o}rte Becher, Christine Klockow, Antje
  Gardebrecht, Christin~M Bennke, Mariette Kassabgy, Sixing Huang, Alexander~J
  Mann, Jost Waldmann, et~al.
\newblock Substrate-controlled succession of marine bacterioplankton
  populations induced by a phytoplankton bloom.
\newblock \emph{Science}, 336\penalty0 (6081):\penalty0 608--611, 2012.

\bibitem[Sun et~al.(2011)Sun, Steindler, Thrash, Halsey, Smith, Carter, Landry,
  and Giovannoni]{sun2011one}
Jing Sun, Laura Steindler, J~Cameron Thrash, Kimberly~H Halsey, Daniel~P Smith,
  Amy~E Carter, Zachary~C Landry, and Stephen~J Giovannoni.
\newblock One carbon metabolism in sar11 pelagic marine bacteria.
\newblock \emph{PloS one}, 6\penalty0 (8):\penalty0 e23973, 2011.

\bibitem[Venceslau et~al.(2010)Venceslau, Lino, and Pereira]{venceslau2010qrc}
Sofia~S Venceslau, Rita~R Lino, and Ines~AC Pereira.
\newblock The qrc membrane complex, related to the alternative complex iii, is
  a menaquinone reductase involved in sulfate respiration.
\newblock \emph{Journal of Biological Chemistry}, 285\penalty0 (30):\penalty0
  22774--22783, 2010.

\bibitem[Tian et~al.(2020)Tian, Ning, He, Zhang, Spencer, Gao, Shi, Wu, Zhang,
  Yang, et~al.]{tian2020small}
Renmao Tian, Daliang Ning, Zhili He, Ping Zhang, Sarah~J Spencer, Shuhong Gao,
  Weiling Shi, Linwei Wu, Ya~Zhang, Yunfeng Yang, et~al.
\newblock Small and mighty: adaptation of superphylum patescibacteria to
  groundwater environment drives their genome simplicity.
\newblock \emph{Microbiome}, 8\penalty0 (1):\penalty0 1--15, 2020.

\bibitem[Grote et~al.(2012)Grote, Thrash, Huggett, Landry, Carini, Giovannoni,
  and Rapp{\'e}]{grote2012streamlining}
Jana Grote, J~Cameron Thrash, Megan~J Huggett, Zachary~C Landry, Paul Carini,
  Stephen~J Giovannoni, and Michael~S Rapp{\'e}.
\newblock Streamlining and core genome conservation among highly divergent
  members of the sar11 clade.
\newblock \emph{MBio}, 3\penalty0 (5):\penalty0 e00252--12, 2012.

\bibitem[Darby et~al.(2007)Darby, Cho, Fuxelius, Westberg, and
  Andersson]{darby2007intracellular}
Alistair~C Darby, Nam-Huyk Cho, Hans-Henrik Fuxelius, Joakim Westberg, and
  Siv~GE Andersson.
\newblock Intracellular pathogens go extreme: genome evolution in the
  rickettsiales.
\newblock \emph{TRENDS in Genetics}, 23\penalty0 (10):\penalty0 511--520, 2007.

\bibitem[Giovannoni et~al.(2014)Giovannoni, Cameron~Thrash, and
  Temperton]{giovannoni2014implications}
Stephen~J Giovannoni, J~Cameron~Thrash, and Ben Temperton.
\newblock Implications of streamlining theory for microbial ecology.
\newblock \emph{The ISME journal}, 8\penalty0 (8):\penalty0 1553--1565, 2014.

\bibitem[Blankenship(2010)]{blankenship2010early}
Robert~E Blankenship.
\newblock Early evolution of photosynthesis.
\newblock \emph{Plant physiology}, 154\penalty0 (2):\penalty0 434--438, 2010.

\bibitem[Ferenci(2016)]{ferenci2016trade}
Thomas Ferenci.
\newblock Trade-off mechanisms shaping the diversity of bacteria.
\newblock \emph{Trends in microbiology}, 24\penalty0 (3):\penalty0 209--223,
  2016.

\bibitem[Lindh and Pinhassi(2018)]{lindh2018sensitivity}
Markus~V Lindh and Jarone Pinhassi.
\newblock Sensitivity of bacterioplankton to environmental disturbance: a
  review of baltic sea field studies and experiments.
\newblock \emph{Frontiers in Marine Science}, 5:\penalty0 361, 2018.

\bibitem[Bertos-Fortis et~al.(2016)Bertos-Fortis, Farnelid, Lindh, Casini,
  Andersson, Pinhassi, and Legrand]{bertos2016unscrambling}
Mireia Bertos-Fortis, Hanna~M Farnelid, Markus~V Lindh, Michele Casini, Agneta
  Andersson, Jarone Pinhassi, and Catherine Legrand.
\newblock Unscrambling cyanobacteria community dynamics related to
  environmental factors.
\newblock \emph{Frontiers in microbiology}, 7:\penalty0 625, 2016.

\bibitem[Bunse et~al.(2019)Bunse, Israelsson, Baltar, Bertos-Fortis,
  Fridolfsson, Legrand, Lindehoff, Lindh, Mart{\'\i}nez-Garc{\'\i}a, and
  Pinhassi]{bunse2019high}
Carina Bunse, Stina Israelsson, Federico Baltar, Mireia Bertos-Fortis, Emil
  Fridolfsson, Catherine Legrand, Elin Lindehoff, Markus~V Lindh, Sandra
  Mart{\'\i}nez-Garc{\'\i}a, and Jarone Pinhassi.
\newblock High frequency multi-year variability in baltic sea microbial
  plankton stocks and activities.
\newblock \emph{Frontiers in microbiology}, 9:\penalty0 3296, 2019.

\bibitem[M{\"u}hlenbruch et~al.(2018)M{\"u}hlenbruch, Grossart, Eigemann, and
  Voss]{muhlenbruch2018mini}
Marco M{\"u}hlenbruch, Hans-Peter Grossart, Falk Eigemann, and Maren Voss.
\newblock Mini-review: Phytoplankton-derived polysaccharides in the marine
  environment and their interactions with heterotrophic bacteria.
\newblock \emph{Environmental microbiology}, 20\penalty0 (8):\penalty0
  2671--2685, 2018.

\bibitem[Ferrer-Gonz{\'a}lez et~al.(2021)Ferrer-Gonz{\'a}lez, Widner,
  Holderman, Glushka, Edison, Kujawinski, and Moran]{ferrer2021resource}
Frank~Xavier Ferrer-Gonz{\'a}lez, Brittany Widner, Nicole~R Holderman, John
  Glushka, Arthur~S Edison, Elizabeth~B Kujawinski, and Mary~Ann Moran.
\newblock Resource partitioning of phytoplankton metabolites that support
  bacterial heterotrophy.
\newblock \emph{The ISME journal}, 15\penalty0 (3):\penalty0 762--773, 2021.

\bibitem[Newton et~al.(2010)Newton, Griffin, Bowles, Meile, Gifford, Givens,
  Howard, King, Oakley, Reisch, et~al.]{newton2010genome}
Ryan~J Newton, Laura~E Griffin, Kathy~M Bowles, Christof Meile, Scott Gifford,
  Carrie~E Givens, Erinn~C Howard, Eric King, Clinton~A Oakley, Chris~R Reisch,
  et~al.
\newblock Genome characteristics of a generalist marine bacterial lineage.
\newblock \emph{The ISME journal}, 4\penalty0 (6):\penalty0 784--798, 2010.

\bibitem[Tsubouchi et~al.(2014)Tsubouchi, Ohta, Haga, Usui, Shimane, Mori,
  Tanizaki, Adachi, Kobayashi, Yukawa, et~al.]{tsubouchi2014thalassospira}
Taishi Tsubouchi, Yukari Ohta, Takuma Haga, Keiko Usui, Yasuhiro Shimane, Kozue
  Mori, Akiko Tanizaki, Akiko Adachi, Kiwa Kobayashi, Kiyotaka Yukawa, et~al.
\newblock Thalassospira alkalitolerans sp. nov. and thalassospira mesophila sp.
  nov., isolated from a decaying bamboo sunken in the marine environment, and
  emended description of the genus thalassospira.
\newblock \emph{International journal of systematic and evolutionary
  microbiology}, 64\penalty0 (Pt\_1):\penalty0 107--115, 2014.

\bibitem[Bergen et~al.(2018)Bergen, Naumann, Herlemann, Gr{\"a}we, Labrenz, and
  J{\"u}rgens]{bergen2018impact}
Benjamin Bergen, Michael Naumann, Daniel~PR Herlemann, Ulf Gr{\"a}we, Matthias
  Labrenz, and Klaus J{\"u}rgens.
\newblock Impact of a major inflow event on the composition and distribution of
  bacterioplankton communities in the baltic sea.
\newblock \emph{Frontiers in Marine Science}, page 383, 2018.

\bibitem[Vetterli et~al.(2015)Vetterli, Hyyti{\"a}inen, Ahjos, Auvinen, Paulin,
  Hietanen, and Leskinen]{vetterli2015seasonal}
Adrien Vetterli, Kirsi Hyyti{\"a}inen, Minttu Ahjos, Petri Auvinen, Lars
  Paulin, Susanna Hietanen, and Elina Leskinen.
\newblock Seasonal patterns of bacterial communities in the coastal brackish
  sediments of the gulf of finland, baltic sea.
\newblock \emph{Estuarine, Coastal and Shelf Science}, 165:\penalty0 86--96,
  2015.

\bibitem[Korneeva et~al.(2015)Korneeva, Pimenov, Krek, Tourova, and
  Bryukhanov]{korneeva2015sulfate}
VA~Korneeva, NV~Pimenov, AV~Krek, TP~Tourova, and AL~Bryukhanov.
\newblock Sulfate-reducing bacterial communities in the water column of the
  gdansk deep (baltic sea).
\newblock \emph{Microbiology}, 84\penalty0 (2):\penalty0 268--277, 2015.

\bibitem[Reissmann et~al.(2009)Reissmann, Burchard, Feistel, Hagen, Lass,
  Mohrholz, Nausch, Umlauf, and Wieczorek]{reissmann2009vertical}
Jan~Hinrich Reissmann, Hans Burchard, Rainer Feistel, Eberhard Hagen,
  Hans~Ulrich Lass, Volker Mohrholz, G{\"u}nther Nausch, Lars Umlauf, and Gunda
  Wieczorek.
\newblock Vertical mixing in the baltic sea and consequences for
  eutrophication--a review.
\newblock \emph{Progress in Oceanography}, 82\penalty0 (1):\penalty0 47--80,
  2009.

\bibitem[Herlemann et~al.(2016)Herlemann, Lundin, Andersson, Labrenz, and
  J{\"u}rgens]{herlemann2016phylogenetic}
Daniel~PR Herlemann, Daniel Lundin, Anders~F Andersson, Matthias Labrenz, and
  Klaus J{\"u}rgens.
\newblock Phylogenetic signals of salinity and season in bacterial community
  composition across the salinity gradient of the baltic sea.
\newblock \emph{Frontiers in Microbiology}, 7:\penalty0 1883, 2016.

\bibitem[Lindh et~al.(2015)Lindh, Sj{\"o}stedt, Andersson, Baltar, Hugerth,
  Lundin, Muthusamy, Legrand, and Pinhassi]{lindh2015disentangling}
Markus~V Lindh, Johanna Sj{\"o}stedt, Anders~F Andersson, Federico Baltar,
  Luisa~W Hugerth, Daniel Lundin, Saraladevi Muthusamy, Catherine Legrand, and
  Jarone Pinhassi.
\newblock Disentangling seasonal bacterioplankton population dynamics by
  high-frequency sampling.
\newblock \emph{Environmental microbiology}, 17\penalty0 (7):\penalty0
  2459--2476, 2015.

\bibitem[Rieck et~al.(2015)Rieck, Herlemann, J{\"u}rgens, and
  Grossart]{rieck2015particle}
Angelika Rieck, Daniel~PR Herlemann, Klaus J{\"u}rgens, and Hans-Peter
  Grossart.
\newblock Particle-associated differ from free-living bacteria in surface
  waters of the baltic sea.
\newblock \emph{Frontiers in Microbiology}, 6:\penalty0 1297, 2015.

\bibitem[Raes et~al.(2022)Raes, Tolman, Desai, Ratten, Zorz, Robicheau, Haider,
  and LaRoche]{raes2022seasonal}
Eric Raes, Jennifer Tolman, Dhwani Desai, Jenni-Marie Ratten, Jackie Zorz,
  Brent~M Robicheau, Diana Haider, and Julie LaRoche.
\newblock Seasonal bacterial niche structures and chemolithoautotrophic
  ecotypes in a north atlantic fjord.
\newblock doi:doi.org/10.21203/rs.3.rs-1469763/v1, 2022.

\bibitem[Pernthaler(2017)]{pernthaler2017competition}
Jakob Pernthaler.
\newblock Competition and niche separation of pelagic bacteria in freshwater
  habitats.
\newblock \emph{Environmental microbiology}, 19\penalty0 (6):\penalty0
  2133--2150, 2017.

\bibitem[Pianka et~al.(2017)Pianka, Vitt, Pelegrin, Fitzgerald, and
  Winemiller]{pianka2017toward}
Eric~R Pianka, Laurie~J Vitt, Nicol{\'a}s Pelegrin, Daniel~B Fitzgerald, and
  Kirk~O Winemiller.
\newblock Toward a periodic table of niches, or exploring the lizard niche
  hypervolume.
\newblock \emph{The American Naturalist}, 190\penalty0 (5):\penalty0 601--616,
  2017.

\bibitem[Gogarten and Townsend(2005)]{gogarten2005horizontal}
J~Peter Gogarten and Jeffrey~P Townsend.
\newblock Horizontal gene transfer, genome innovation and evolution.
\newblock \emph{Nature Reviews Microbiology}, 3\penalty0 (9):\penalty0
  679--687, 2005.

\bibitem[Ward et~al.(2017)Ward, Yung, Davis, Blinebry, Williams, Johnson, and
  Hunt]{ward2017annual}
Christopher~S Ward, Cheuk-Man Yung, Katherine~M Davis, Sara~K Blinebry,
  Tiffany~C Williams, Zackary~I Johnson, and Dana~E Hunt.
\newblock Annual community patterns are driven by seasonal switching between
  closely related marine bacteria.
\newblock \emph{The ISME journal}, 11\penalty0 (6):\penalty0 1412--1422, 2017.

\bibitem[Seidel et~al.(2017)Seidel, Manecki, Herlemann, Deutsch, Schulz-Bull,
  J{\"u}rgens, and Dittmar]{seidel2017composition}
Michael Seidel, Marcus Manecki, Daniel~PR Herlemann, Barbara Deutsch, Detlef
  Schulz-Bull, Klaus J{\"u}rgens, and Thorsten Dittmar.
\newblock Composition and transformation of dissolved organic matter in the
  baltic sea.
\newblock \emph{Frontiers in Earth Science}, 5:\penalty0 31, 2017.

\bibitem[Alneberg et~al.(2020)Alneberg, Bennke, Beier, Bunse, Quince,
  Ininbergs, Riemann, Ekman, J{\"u}rgens, Labrenz,
  et~al.]{alneberg2020ecosystem}
Johannes Alneberg, Christin Bennke, Sara Beier, Carina Bunse, Christopher
  Quince, Karolina Ininbergs, Lasse Riemann, Martin Ekman, Klaus J{\"u}rgens,
  Matthias Labrenz, et~al.
\newblock Ecosystem-wide metagenomic binning enables prediction of ecological
  niches from genomes.
\newblock \emph{Communications biology}, 3\penalty0 (1):\penalty0 1--10, 2020.

\bibitem[Gowda et~al.(2022)Gowda, Ping, Mani, and Kuehn]{gowda2022genomic}
Karna Gowda, Derek Ping, Madhav Mani, and Seppe Kuehn.
\newblock Genomic structure predicts metabolite dynamics in microbial
  communities.
\newblock \emph{Cell}, 2022.

\bibitem[Garc{\'\i}a-Garc{\'\i}a et~al.(2019)Garc{\'\i}a-Garc{\'\i}a, Tamames,
  Linz, Pedr{\'o}s-Ali{\'o}, and Puente-S{\'a}nchez]{garcia2019microdiversity}
Natalia Garc{\'\i}a-Garc{\'\i}a, Javier Tamames, Alexandra~M Linz, Carlos
  Pedr{\'o}s-Ali{\'o}, and Fernando Puente-S{\'a}nchez.
\newblock Microdiversity ensures the maintenance of functional microbial
  communities under changing environmental conditions.
\newblock \emph{The ISME journal}, 13\penalty0 (12):\penalty0 2969--2983, 2019.

\bibitem[Zhu et~al.(2015)Zhu, Sunagawa, Mende, and Bork]{zhu2015inter}
Ana Zhu, Shinichi Sunagawa, Daniel~R Mende, and Peer Bork.
\newblock Inter-individual differences in the gene content of human gut
  bacterial species.
\newblock \emph{Genome biology}, 16\penalty0 (1):\penalty0 1--13, 2015.

\bibitem[Forcone et~al.(2021)Forcone, Coutinho, Cavalcanti, and
  Silveira]{forcone2021prophage}
Kathryn Forcone, Felipe~H Coutinho, Giselle~S Cavalcanti, and Cynthia~B
  Silveira.
\newblock Prophage genomics and ecology in the family rhodobacteraceae.
\newblock \emph{Microorganisms}, 9\penalty0 (6):\penalty0 1115, 2021.

\bibitem[Bostr{\"o}m et~al.(2004)Bostr{\"o}m, Simu, Hagstr{\"o}m, and
  Riemann]{bostrom2004optimization}
Kj{\"a}rstin~H Bostr{\"o}m, Karin Simu, {\AA}ke Hagstr{\"o}m, and Lasse
  Riemann.
\newblock Optimization of dna extraction for quantitative marine
  bacterioplankton community analysis.
\newblock \emph{Limnology and Oceanography: Methods}, 2\penalty0 (11):\penalty0
  365--373, 2004.

\bibitem[Bunse et~al.(2016)Bunse, Bertos-Fortis, Sassenhagen, Sildever,
  Sj{\"o}qvist, Godhe, Gross, Kremp, Lips, Lundholm, et~al.]{bunse2016spatio}
Carina Bunse, Mireia Bertos-Fortis, Ingrid Sassenhagen, Sirje Sildever, Conny
  Sj{\"o}qvist, Anna Godhe, Susanna Gross, Anke Kremp, Inga Lips, Nina
  Lundholm, et~al.
\newblock Spatio-temporal interdependence of bacteria and phytoplankton during
  a baltic sea spring bloom.
\newblock \emph{Frontiers in microbiology}, 7:\penalty0 517, 2016.

\bibitem[Herlemann et~al.(2011)Herlemann, Labrenz, J{\"u}rgens, Bertilsson,
  Waniek, and Andersson]{herlemann2011transitions}
Daniel~PR Herlemann, Matthias Labrenz, Klaus J{\"u}rgens, Stefan Bertilsson,
  Joanna~J Waniek, and Anders~F Andersson.
\newblock Transitions in bacterial communities along the 2000 km salinity
  gradient of the baltic sea.
\newblock \emph{The ISME journal}, 5\penalty0 (10):\penalty0 1571--1579, 2011.

\bibitem[Hugerth et~al.(2014)Hugerth, Wefer, Lundin, Jakobsson, Lindberg,
  Rodin, Engstrand, and Andersson]{hugerth2014degeprime}
Luisa~W Hugerth, Hugo~A Wefer, Sverker Lundin, Hedvig~E Jakobsson, Mathilda
  Lindberg, Sandra Rodin, Lars Engstrand, and Anders~F Andersson.
\newblock Degeprime, a program for degenerate primer design for
  broad-taxonomic-range pcr in microbial ecology studies.
\newblock \emph{Applied and environmental microbiology}, 80\penalty0
  (16):\penalty0 5116--5123, 2014.

\bibitem[Ewels et~al.(2020)Ewels, Peltzer, Fillinger, Patel, Alneberg, Wilm,
  Garcia, Di~Tommaso, and Nahnsen]{ewels2020nf}
Philip~A Ewels, Alexander Peltzer, Sven Fillinger, Harshil Patel, Johannes
  Alneberg, Andreas Wilm, Maxime~Ulysse Garcia, Paolo Di~Tommaso, and Sven
  Nahnsen.
\newblock The nf-core framework for community-curated bioinformatics pipelines.
\newblock \emph{Nature biotechnology}, 38\penalty0 (3):\penalty0 276--278,
  2020.

\bibitem[Straub et~al.(2020)Straub, Blackwell, Langarica-Fuentes, Peltzer,
  Nahnsen, and Kleindienst]{straub2020interpretations}
Daniel Straub, Nia Blackwell, Adrian Langarica-Fuentes, Alexander Peltzer, Sven
  Nahnsen, and Sara Kleindienst.
\newblock Interpretations of environmental microbial community studies are
  biased by the selected 16s rrna (gene) amplicon sequencing pipeline.
\newblock \emph{Frontiers in Microbiology}, 11:\penalty0 550420, 2020.

\bibitem[Callahan et~al.(2016)Callahan, McMurdie, Rosen, Han, Johnson, and
  Holmes]{callahan2016dada2}
Benjamin~J Callahan, Paul~J McMurdie, Michael~J Rosen, Andrew~W Han, Amy Jo~A
  Johnson, and Susan~P Holmes.
\newblock Dada2: High-resolution sample inference from illumina amplicon data.
\newblock \emph{Nature methods}, 13\penalty0 (7):\penalty0 581--583, 2016.

\bibitem[Bolyen et~al.(2019)Bolyen, Rideout, Dillon, Bokulich, Abnet,
  Al-Ghalith, Alexander, Alm, Arumugam, Asnicar,
  et~al.]{bolyen2019reproducible}
Evan Bolyen, Jai~Ram Rideout, Matthew~R Dillon, Nicholas~A Bokulich,
  Christian~C Abnet, Gabriel~A Al-Ghalith, Harriet Alexander, Eric~J Alm,
  Manimozhiyan Arumugam, Francesco Asnicar, et~al.
\newblock Reproducible, interactive, scalable and extensible microbiome data
  science using qiime 2.
\newblock \emph{Nature biotechnology}, 37\penalty0 (8):\penalty0 852--857,
  2019.

\bibitem[Seemann(2014)]{seemann2014prokka}
Torsten Seemann.
\newblock Prokka: rapid prokaryotic genome annotation.
\newblock \emph{Bioinformatics}, 30\penalty0 (14):\penalty0 2068--2069, 2014.

\bibitem[Subramanian et~al.(2005)Subramanian, Tamayo, Mootha, Mukherjee, Ebert,
  Gillette, Paulovich, Pomeroy, Golub, Lander, et~al.]{subramanian2005gene}
Aravind Subramanian, Pablo Tamayo, Vamsi~K Mootha, Sayan Mukherjee, Benjamin~L
  Ebert, Michael~A Gillette, Amanda Paulovich, Scott~L Pomeroy, Todd~R Golub,
  Eric~S Lander, et~al.
\newblock Gene set enrichment analysis: a knowledge-based approach for
  interpreting genome-wide expression profiles.
\newblock \emph{Proceedings of the National Academy of Sciences}, 102\penalty0
  (43):\penalty0 15545--15550, 2005.

\bibitem[{R Core Team}(2020)]{R}
{R Core Team}.
\newblock \emph{R: A Language and Environment for Statistical Computing}.
\newblock R Foundation for Statistical Computing, Vienna, Austria, 2020.
\newblock URL \url{https://www.R-project.org/}.

\bibitem[Korotkevich et~al.(2019)Korotkevich, Sukhov, and Sergushichev]{fgsea}
Gennady Korotkevich, Vladimir Sukhov, and Alexey Sergushichev.
\newblock Fast gene set enrichment analysis.
\newblock \emph{bioRxiv}, 2019.
\newblock \doi{10.1101/060012}.
\newblock URL \url{http://biorxiv.org/content/early/2016/06/20/060012}.

\bibitem[Benjamini and Hochberg(1995)]{benjamini1995controlling}
Yoav Benjamini and Yosef Hochberg.
\newblock Controlling the false discovery rate: a practical and powerful
  approach to multiple testing.
\newblock \emph{Journal of the Royal statistical society: series B
  (Methodological)}, 57\penalty0 (1):\penalty0 289--300, 1995.

\bibitem[Bezanson et~al.(2017)Bezanson, Edelman, Karpinski, and
  Shah]{Julia-2017}
Jeff Bezanson, Alan Edelman, Stefan Karpinski, and Viral~B Shah.
\newblock Julia: A fresh approach to numerical computing.
\newblock \emph{SIAM {R}eview}, 59\penalty0 (1):\penalty0 65--98, 2017.
\newblock \doi{10.1137/141000671}.
\newblock URL \url{https://epubs.siam.org/doi/10.1137/141000671}.

\bibitem[Wickham(2016)]{ggplot2}
Hadley Wickham.
\newblock \emph{ggplot2: Elegant Graphics for Data Analysis}.
\newblock Springer-Verlag New York, 2016.
\newblock ISBN 978-3-319-24277-4.
\newblock URL \url{https://ggplot2.tidyverse.org}.

\bibitem[Wickham et~al.(2019)Wickham, Averick, Bryan, Chang, McGowan,
  François, Grolemund, Hayes, Henry, Hester, Kuhn, Pedersen, Miller, Bache,
  Müller, Ooms, Robinson, Seidel, Spinu, Takahashi, Vaughan, Wilke, Woo, and
  Yutani]{tidyverse}
Hadley Wickham, Mara Averick, Jennifer Bryan, Winston Chang, Lucy~D'Agostino
  McGowan, Romain François, Garrett Grolemund, Alex Hayes, Lionel Henry, Jim
  Hester, Max Kuhn, Thomas~Lin Pedersen, Evan Miller, Stephan~Milton Bache,
  Kirill Müller, Jeroen Ooms, David Robinson, Dana~Paige Seidel, Vitalie
  Spinu, Kohske Takahashi, Davis Vaughan, Claus Wilke, Kara Woo, and Hiroaki
  Yutani.
\newblock Welcome to the {tidyverse}.
\newblock \emph{Journal of Open Source Software}, 4\penalty0 (43):\penalty0
  1686, 2019.
\newblock \doi{10.21105/joss.01686}.

\bibitem[Wickham(2011)]{plyr}
Hadley Wickham.
\newblock The split-apply-combine strategy for data analysis.
\newblock \emph{Journal of Statistical Software}, 40\penalty0 (1):\penalty0
  1--29, 2011.
\newblock URL \url{https://www.jstatsoft.org/v40/i01/}.

\bibitem[Xu et~al.(2021)Xu, Chen, Feng, Zhan, Zhou, and Yu]{ggbreak}
Shuangbin Xu, Meijun Chen, Tingze Feng, Li~Zhan, Lang Zhou, and Guangchuang Yu.
\newblock Use ggbreak to effectively utilize plotting space to deal with large
  datasets and outliers.
\newblock \emph{Frontiers in Genetics}, 12:\penalty0 774846, 2021.
\newblock \doi{10.3389/fgene.2021.774846}.

\bibitem[Kassambara(2020)]{ggpubr}
Alboukadel Kassambara.
\newblock \emph{ggpubr: 'ggplot2' Based Publication Ready Plots}, 2020.
\newblock URL \url{https://rpkgs.datanovia.com/ggpubr/}.
\newblock R package version 0.4.0.999.

\bibitem[Hahsler et~al.(2008)Hahsler, Hornik, and Buchta]{seriation}
Michael Hahsler, Kurt Hornik, and Christian Buchta.
\newblock Getting things in order: An introduction to the r package seriation.
\newblock \emph{Journal of Statistical Software}, 25\penalty0 (3):\penalty0
  1--34, March 2008.
\newblock ISSN 1548-7660.
\newblock \doi{10.18637/jss.v025.i03}.

\bibitem[Wickham(2007)]{reshape}
Hadley Wickham.
\newblock Reshaping data with the reshape package.
\newblock \emph{Journal of Statistical Software}, 21\penalty0 (12), 2007.
\newblock URL \url{http://www.jstatsoft.org/v21/i12/paper}.

\end{thebibliography}


\end{document}


\begin{flushleft}
{\Large
\textbf\newline{Supplementary material: Quantification of metabolic niche occupancy dynamics in a Baltic Sea bacterial community}
}
\newline
\\
Jana C. Massing\textsuperscript{1,2,3*},
Ashkaan Fahimipour\textsuperscript{4},
Carina Bunse\textsuperscript{1,5},
Jarone Pinhassi\textsuperscript{6},
Thilo Gross\textsuperscript{1,2,3}
\\
\bigskip
{1} Helmholtz Institute for Functional Marine Biodiversity at the University of Oldenburg (HIFMB), Oldenburg, Germany
\\
{2}  Helmholtz Centre for Marine and Polar Research, Alfred-Wegener-Institute, Bremerhaven, Germany
\\
{3} Institute for Chemistry and Biology of the Marine Environment (ICBM), Carl-von-Ossietzky University, Oldenburg, Germany
\\
{4} Institute of Marine Sciences, University of California, Santa Cruz, CA, USA
\\
{5} Department of Marine Sciences, University of Gothenburg, Gothenburg, Sweden 
\\
{6} Centre for Ecology and Evolution in Microbial Model Systems - EEMiS, Linnaeus University, Kalmar, Sweden
\\
\bigskip
* jana.massing@hifmb.de

\section*{Supplementary Tables}


{\tiny
\begin{longtable}{|l|l|l|l|l|l|}
    \caption{\normalsize Genomes that map to the 100 most abundant ASVs obtained from amplicon sequencing data in terms of relative mean abundance over the whole sampling period. Genome, taxonomic information, mean and maximum abundance over the whole sampling period are provided.}
    \\
    \hline
        \textbf{Genome}  & \textbf{Class} & \textbf{Family}& \textbf{Species}  & \textbf{Mean abundance}  & \textbf{Max abundance}  \\ \hline
        GCF\_003011885.1 & Cyanobacteriia & Cyanobiaceae & Cyanobium\_A usitatum & 0.1150 & 0.7222 \\ \hline
        GCF\_002252665.1 & Cyanobacteriia & Cyanobiaceae & Cyanobium\_A sp002252665 & 0.0533 & 0.2779 \\ \hline
        GCA\_003569125.1 & Acidimicrobiia & Ilumatobacteraceae & BACL27 sp003569125 & 0.0488 & 0.3656 \\ \hline
        GCA\_001593825.1 & Cyanobacteriia & Nostocaceae & Aphanizomenon\_B flosaquae & 0.0400 & 0.5437 \\ \hline
        GCF\_000173115.1 & Bacteroidia & Flavobacteriaceae & MAG-120531 sp000173115 & 0.0397 & 0.5443 \\ \hline
        GCA\_002358295.1 & Gammaproteobacteria & D2472 & D2472 sp002358345 & 0.0287 & 0.2754 \\
\hline 
\end{longtable}}


{\tiny
\begin{longtable}{|l|l|l|l|l|}
    \caption{Species that map to the 100 ASVs scoring most negative values in variable~1.}\\     \hline
    \label{Var1_species}
        \textbf{Species}  & \textbf{Value} & \textbf{Genome} & \textbf{Class} & \textbf{Family} \\ \hline
         Salmonella enterica  & -0.09769 & GCA\_007094035.1 & Gammaproteobacteria & Enterobacteriaceae \\ \hline
         Salmonella enterica  & -0.09769 & GCA\_007093895.1 & Gammaproteobacteria & Enterobacteriaceae \\ \hline
         Salmonella enterica  & -0.09763 & GCA\_007093765.1 & Gammaproteobacteria & Enterobacteriaceae \\ \hline
         Salmonella enterica  & -0.09756 & GCF\_003548795.1 & Gammaproteobacteria & Enterobacteriaceae \\ \hline
         Cronobacter sakazakii  & -0.09754 & GCF\_002094495.1 & Gammaproteobacteria & Enterobacteriaceae \\ \hline
         Cronobacter sakazakii  & -0.09754 & GCF\_002094665.1 & Gammaproteobacteria & Enterobacteriaceae \\
\hline
\end{longtable}}


{\footnotesize
\begin{longtable}{|l|l|l|}
    \caption{Top 100 over-represented annotated genes in the genomes of the taxa that receive the most negative entries on variable 1. 
    The \textit{NES} and \textit{FDR-Adj. P} columns show the normalized 'Enrichment score' and FDR-adjusted \cite{benjamini1995controlling} \textit{P}-value from the enrichment analysis \cite{subramanian2005gene}.}\\
    \hline
    \textbf{Gene} & \textbf{FDR-Adj. P} & \textbf{NES} \\ \hline
    Major outer membrane lipoprotein Lpp 1 & -11.844 & 0.00025 \\ \hline
        Phage shock protein G & -11.798 & 0.00025 \\ \hline
        Primosomal replication protein N'' & -11.753 & 0.00025 \\ \hline
        DNA damage-inducible protein I & -11.694 & 0.00025 \\ \hline
        Outer membrane porin C & -11.614 & 0.00025 \\ \hline
        Chaperone protein YcdY & -11.562 & 0.00025 \\
\hline
\end{longtable} }

{\footnotesize
\begin{longtable}{|l|l|l|}
    \caption{Top 100 over-represented annotated genes in the genomes of the taxa that receive the most positive entries on variable 2. The column descriptions are provided with Supplementary Table 3.}\\
    \hline
        \textbf{Gene} & \textbf{FDR-Adj. P} & \textbf{NES} \\
    \hline
        Outer-membrane lipoprotein LolB & 15.094 & 0.0004 \\ \hline
        DNA-binding protein Fis & 15.02 & 0.0004 \\ \hline
        LPS-assembly lipoprotein LptE & 15.004 & 0.0004 \\ \hline
        Ammonia monooxygenase gamma subunit & 14.873 & 0.0004 \\ \hline
        Thiol:disulfide interchange protein DsbA & 14.848 & 0.0004 \\ \hline
        DNA polymerase III subunit delta & 14.719 & 0.0004 \\
\hline
\end{longtable} }

{\footnotesize
\begin{longtable}{|l|l|l|}
    \caption{Top 100 over-represented annotated genes in the genomes of the taxa that receive the most positive entries on variable 3. The column descriptions are provided with Supplementary Table 3.}\\
    \hline
        \textbf{Gene} & \textbf{FDR-Adj. P} & \textbf{NES} \\
    \hline
    GTP pyrophosphokinase rsh & 12.285 & 0.00054 \\ \hline
        putative peptidoglycan D,D-transpeptidase FtsI & 12.172 & 0.00054 \\ \hline
        Chromosome-partitioning protein ParB & 12.152 & 0.00054 \\ \hline
        5-aminolevulinate synthase & 12.02 & 0.00054 \\ \hline
        FtsZ-localized protein C & 11.923 & 0.00054 \\ \hline
        6,7-dimethyl-8-ribityllumazine synthase 1 & 11.812 & 0.00054 \\
\hline
\end{longtable} }


{\footnotesize
\begin{longtable}{|l|l|l|}
    \caption{Top 100 over-represented annotated genes in the genomes of the taxa that receive the most negative entries on variable 4. The column descriptions are provided with Supplementary Table 3.}\\
    \hline
        \textbf{Gene} & \textbf{FDR-Adj. P} & \textbf{NES} \\ \hline
        Cytochrome f & -6.889 & 0.00031 \\ \hline
        Photosystem II manganese-stabilizing polypeptide & -6.876 & 0.00031 \\ \hline
        Protein Thf1 & -6.874 & 0.00031 \\ \hline
        Photosystem II CP47 reaction center protein & -6.859 & 0.00031 \\ \hline
        NAD(P)H-quinone oxidoreductase subunit N & -6.857 & 0.00031 \\ \hline
        Photosystem I assembly protein Ycf4 & -6.857 & 0.00031 \\
\hline
\end{longtable}}


{\footnotesize
\begin{longtable}{|l|l|l|}
    \caption{Top 100 over-represented annotated genes in the genomes of the taxa that receive the most negative entries on variable 14. 
    The column descriptions are provided with Supplementary Table 3.}\\
    \hline
        \textbf{Gene} & \textbf{FDR-Adj. P} & \textbf{NES} \\
    \hline
        Thymidylate synthase 1 & -5.522 & 0.00055 \\ \hline
        DNA-binding protein Bv3F & -5.446 & 0.00055 \\ \hline
        mupirocin-resistant isoleucine--tRNA ligase MupA & -5.294 & 0.00055 \\ \hline
        Cobalt-precorrin-7 C(5)-methyltransferase & -5.114 & 0.00055 \\ \hline
        Cobalt-zinc-cadmium resistance protein CzcI & -5.103 & 0.00055 \\ \hline
        IS200/IS605 family transposase ISCth10 & -4.897 & 0.00055 \\
\hline
\end{longtable}}

{\footnotesize
\begin{longtable}{|l|l|l|}
    \caption{Top 100 over-represented annotated genes in the genomes of the taxa that receive the most negative entries on variable 38. 
    The column descriptions are provided with Supplementary Table 3.}\\
    \hline
        \textbf{Gene} & \textbf{FDR-Adj. P} & \textbf{NES} \\
    \hline
        Bifunctional protein MdtA & -5.039 & 0.00064 \\ \hline
        Flagellar assembly protein FliX & -4.91 & 0.00064 \\ \hline
        Presqualene diphosphate synthase & -4.54 & 0.00064 \\ \hline
        mupirocin-resistant isoleucine--tRNA ligase MupA & -4.537 & 0.00064 \\ \hline
        Formyltransferase/hydrolase complex subunit D & -4.476 & 0.00064 \\ \hline
        Formyltransferase/hydrolase complex Fhc subunit A & -4.455 & 0.00064 \\
\hline
\end{longtable}}

{\footnotesize
\begin{longtable}{|l|l|l|}
    \caption{Top 100 over-represented annotated genes in the genomes of the taxa that receive the most positive entries on variable 43. 
    The column descriptions are provided with Supplementary Table 3.}\\
    \hline
        \textbf{Gene} & \textbf{FDR-Adj. P} & \textbf{NES} \\
    \hline
        Sirohydrochlorin cobaltochelatase CbiKP & 5.05 & 0.00064 \\ \hline
        Bifunctional protein MdtA & 4.982 & 0.00064 \\ \hline
        Cytochrome c-L & 4.843 & 0.00064 \\ \hline
        Methanol dehydrogenase [cytochrome c] subunit 2 & 4.798 & 0.00064 \\ \hline
        Carbon monoxide dehydrogenase 1 & 4.71 & 0.00064 \\ \hline
        NAD(+)--dinitrogen-reductase ADP-D-ribosyltransferase & 4.631 & 0.00064 \\
\hline
\end{longtable}}


\section*{Supplementary Figures}

\begin{figure} [H]
    \centering
    \includegraphics[width=\textwidth]{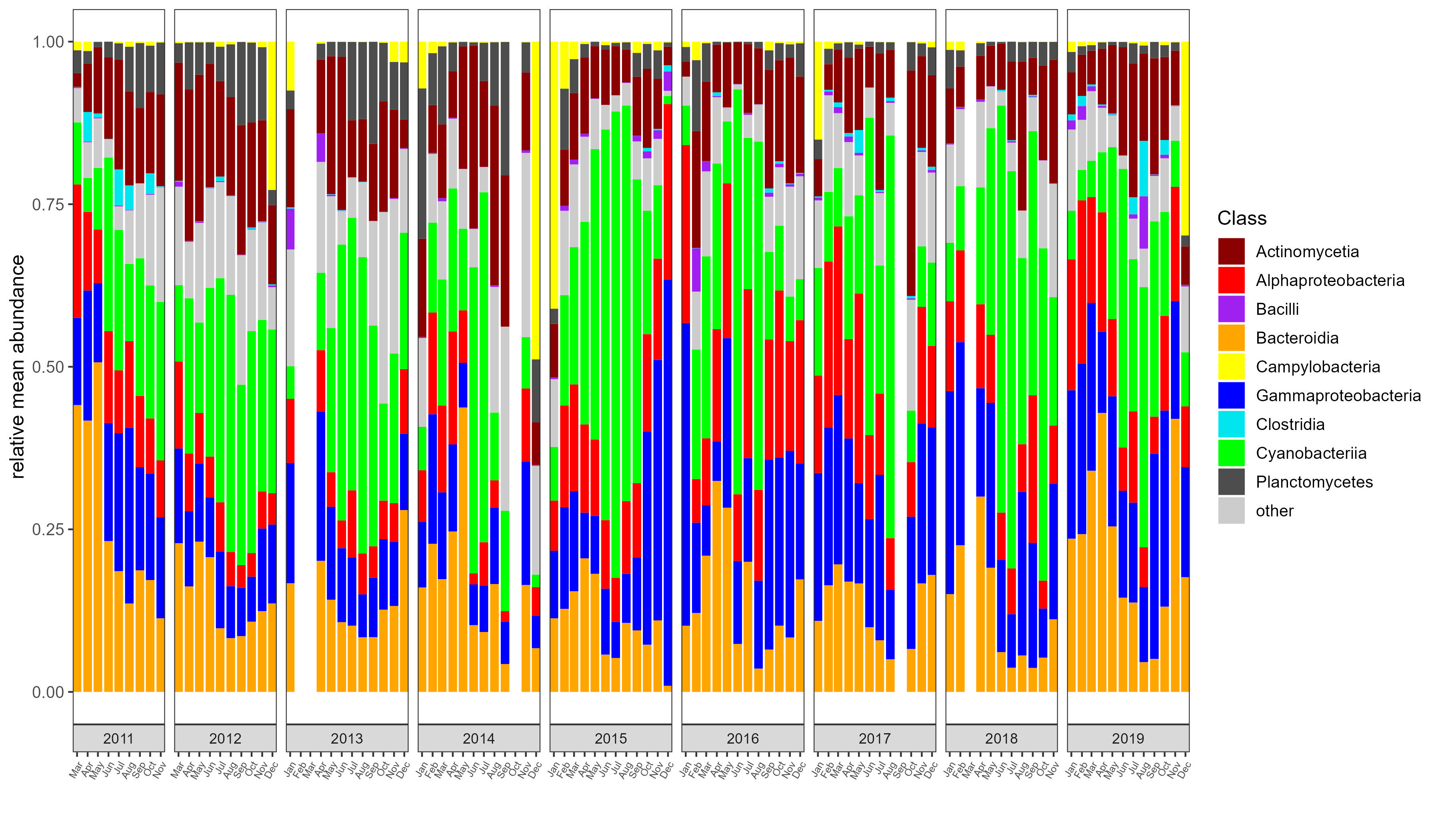}
    \caption{Relative mean abundances of classes that map to the genomes obtained from amplicon sequencing data over the whole sampling period. Taxonomic classes are color-coded.}
\end{figure}


\begin{figure} [H]
    \centering
    \includegraphics[width=\textwidth]{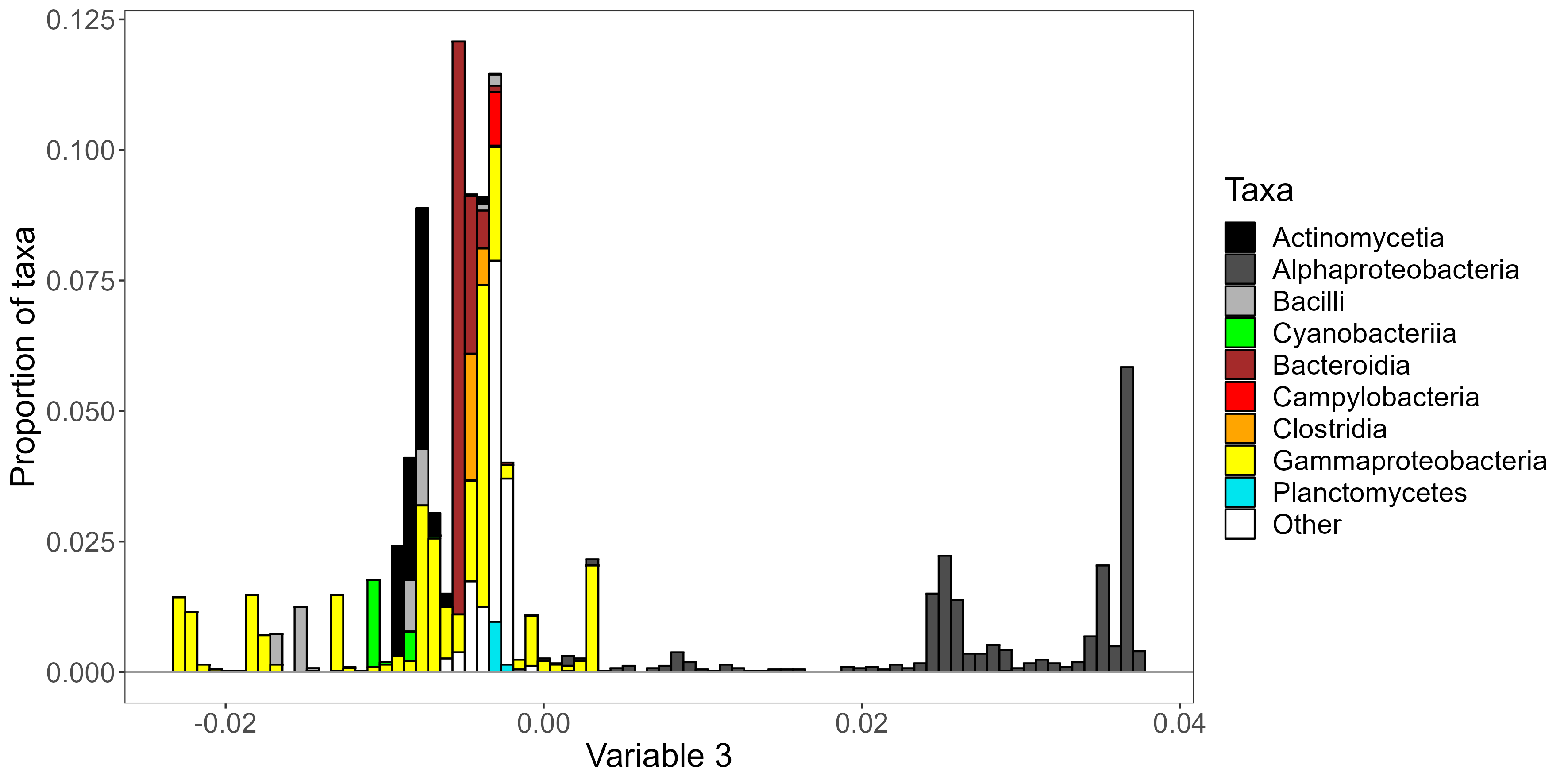}
    \caption{The ordering of taxa defined by variable 3 entries, from negative to positive (left to right). The taxonomic compositions corresponding to variable entries are shown for each of 80 equally spaced bins.}
\end{figure}

\begin{figure} [H]
    \centering
    \includegraphics[width=\textwidth]{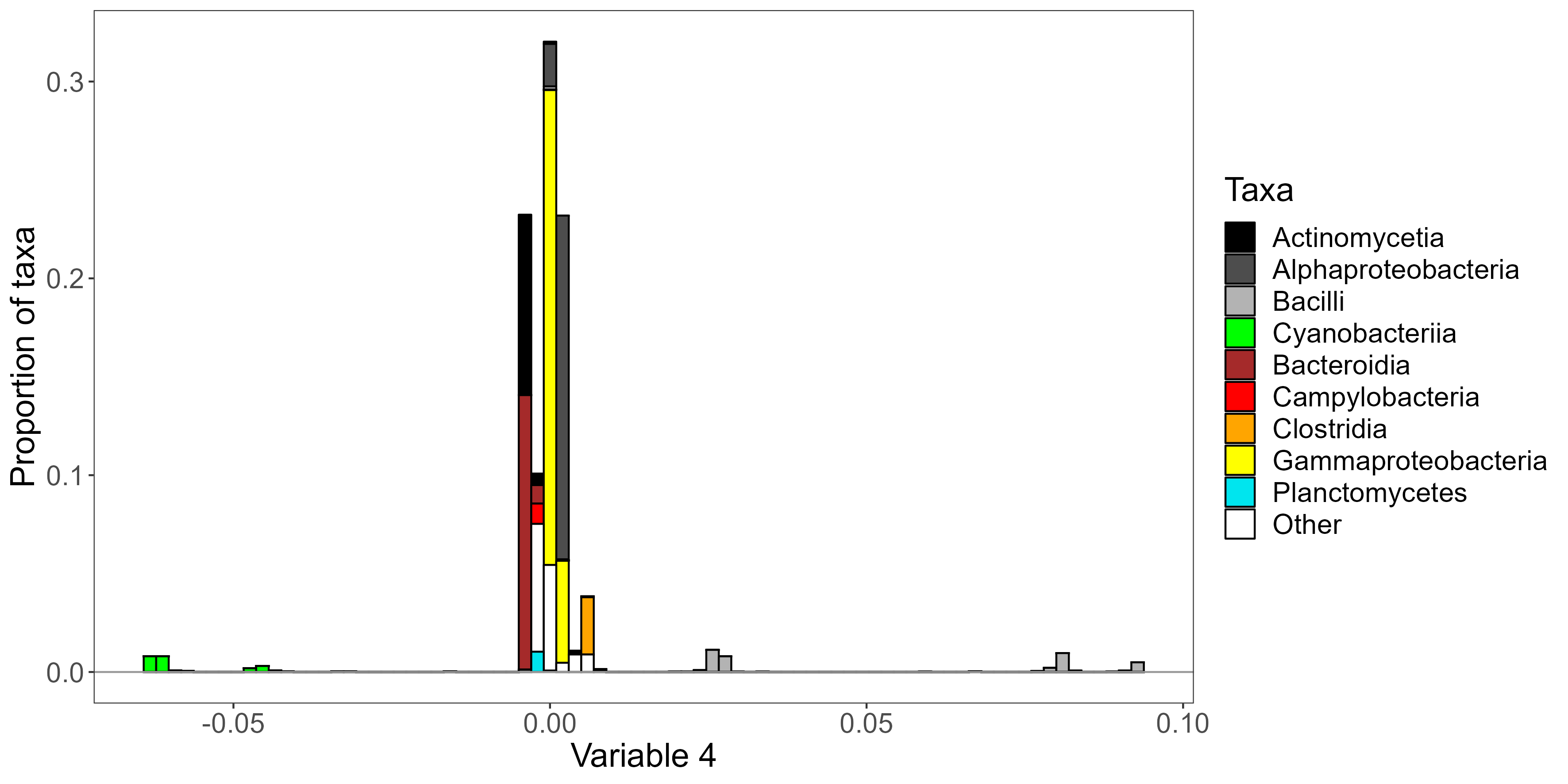}
    \caption{The ordering of taxa defined by variable 4 entries, from negative to positive (left to right). The taxonomic compositions corresponding to variable entries are shown for each of 80 equally spaced bins.}
\end{figure}

\begin{figure} [H]
    \centering
    \includegraphics[width=\textwidth]{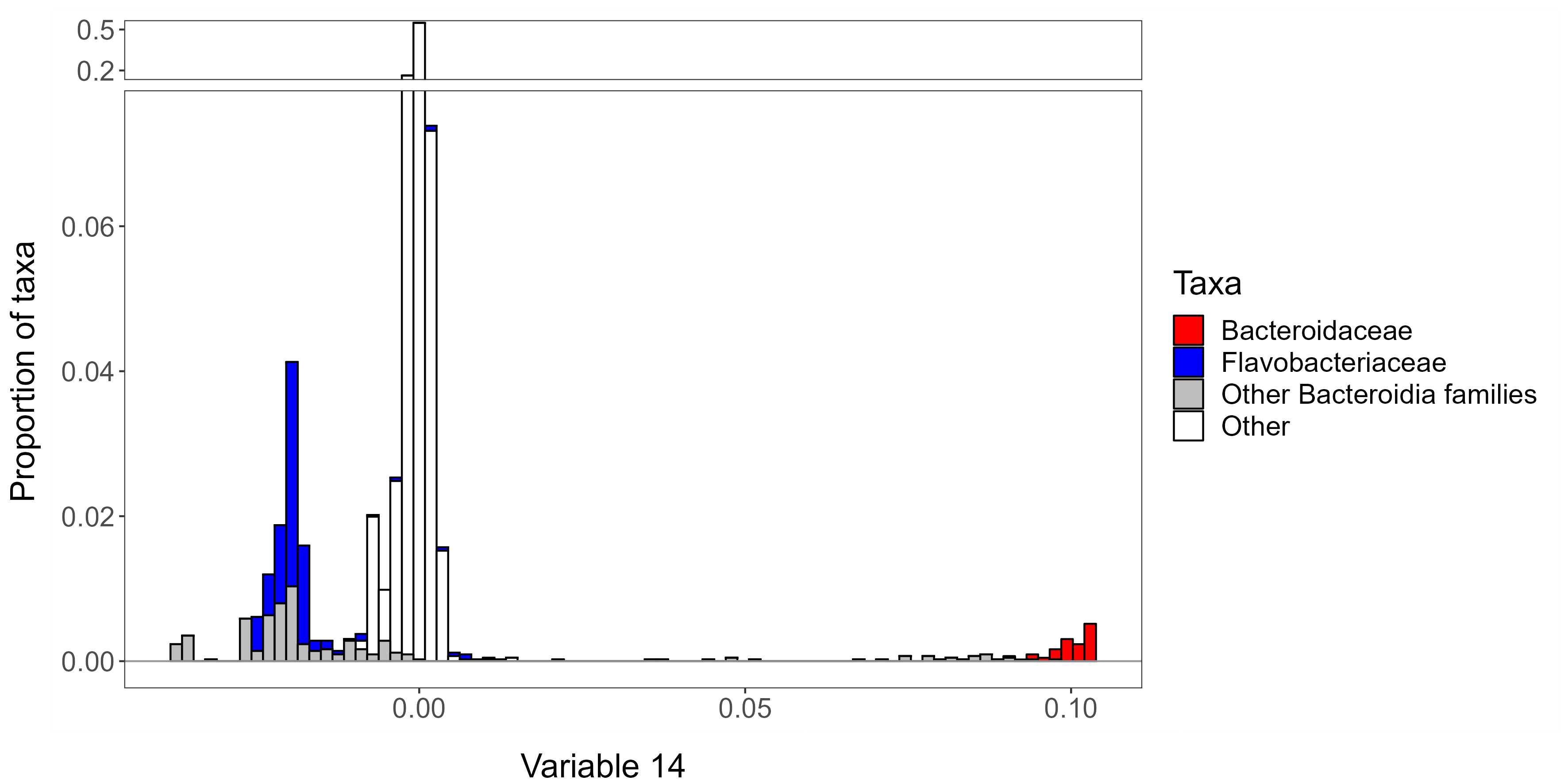}
    \caption{The ordering of taxa defined by variable 14 entries, from negative to positive (left to right). The taxonomic compositions corresponding to variable entries are shown for each of 80 equally spaced bins. Families belonging to the phylum Bacteroidota are color-coded.}
\end{figure}

\begin{figure} [H]
    \centering
    \includegraphics[width=\textwidth]{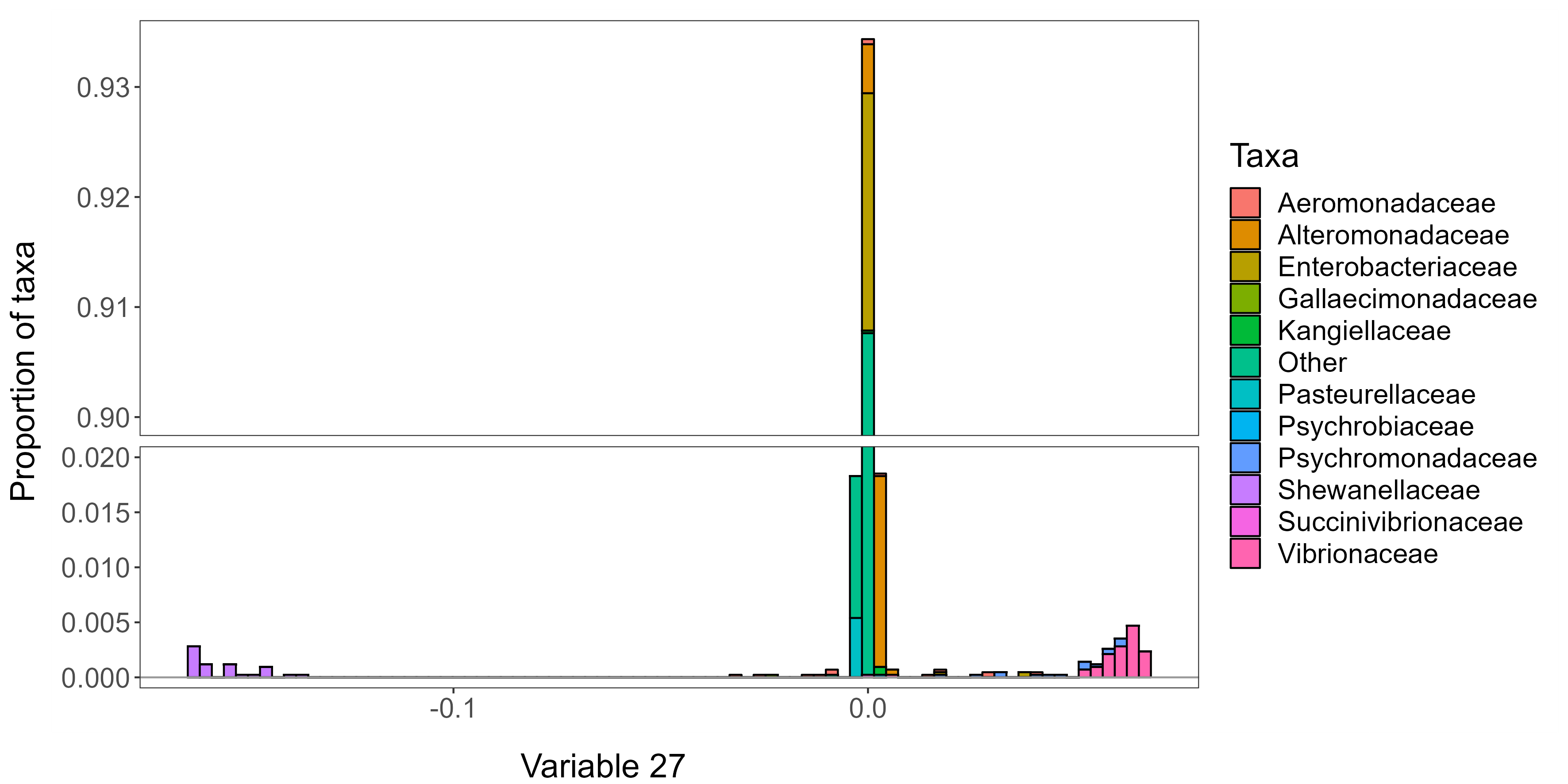}
    \caption{The ordering of taxa defined by variable 27 entries, from negative to positive (left to right). The taxonomic compositions corresponding to variable entries are shown for each of 80 equally spaced bins. Families belonging to the Enterobacterales are color-coded.}
\end{figure}

\begin{figure} [H]
    \centering
    \includegraphics[width=\textwidth]{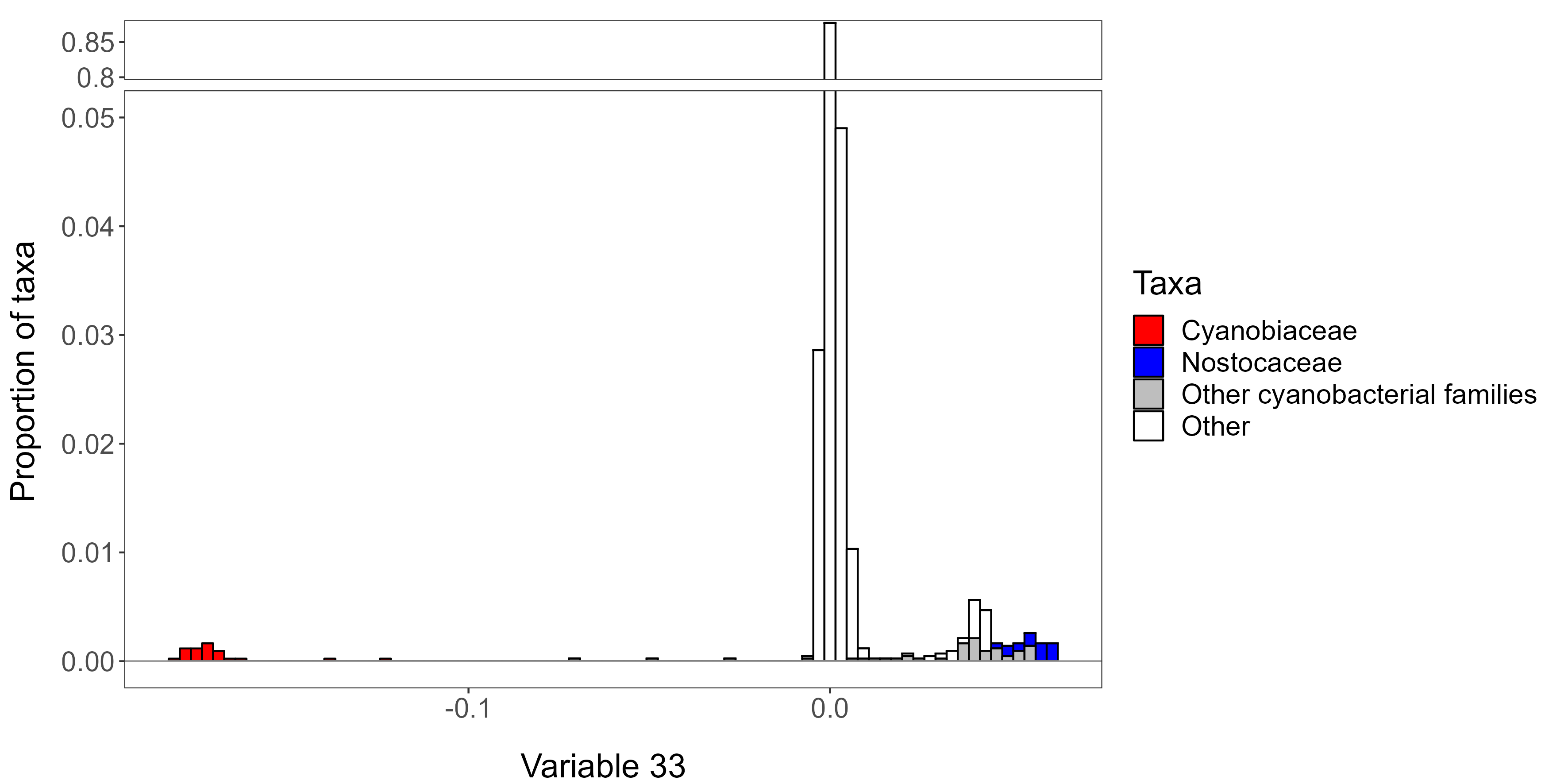}
    \caption{The ordering of taxa defined by variable 33 entries, from negative to positive (left to right). The taxonomic compositions corresponding to variable entries are shown for each of 80 equally spaced bins. Families belonging to the class of Cyanobacteriia are color-coded.}
\end{figure}

\begin{figure} [H]
    \centering
    \includegraphics[width=\textwidth]{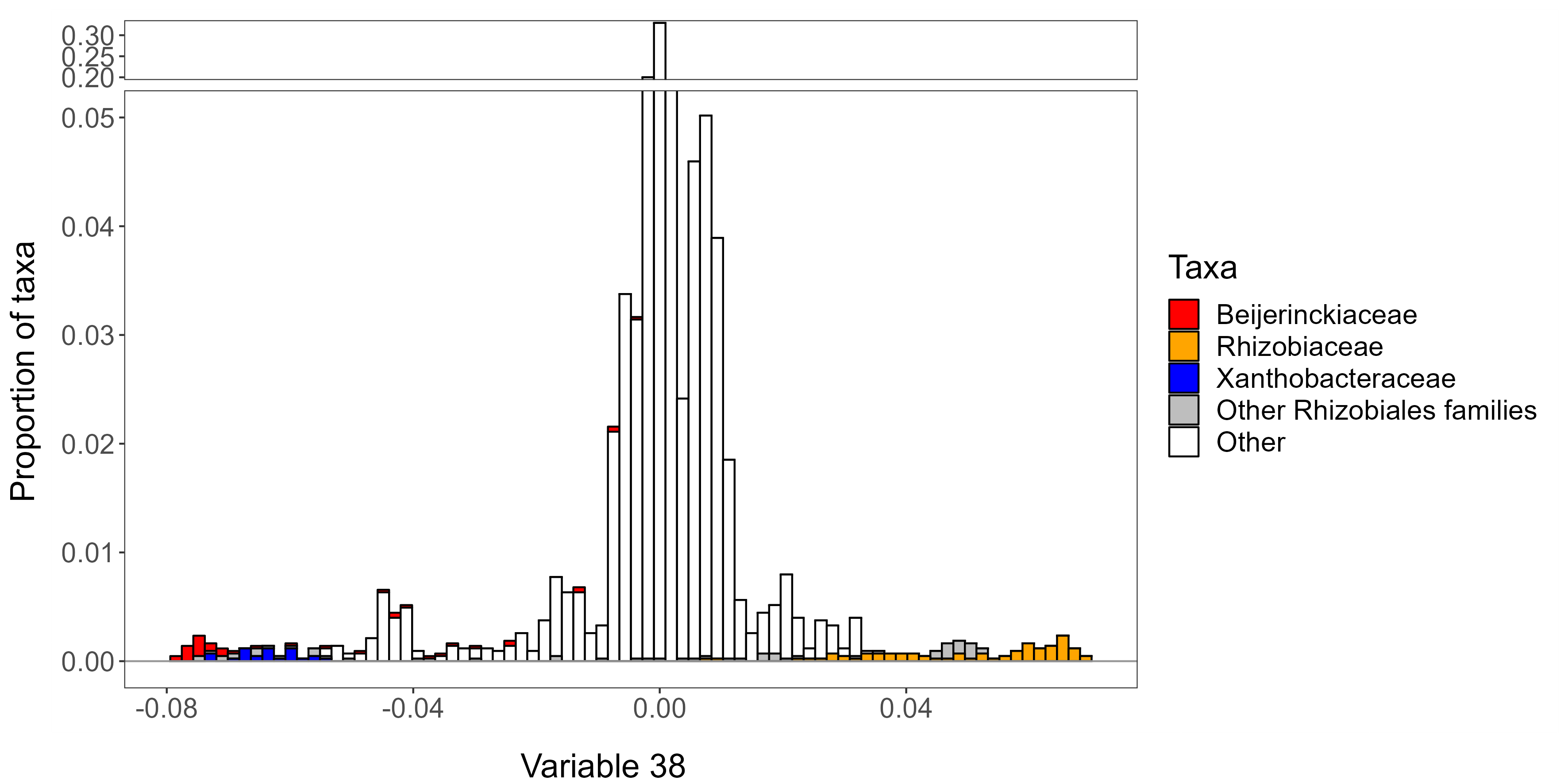}
    \caption{The ordering of taxa defined by variable 38 entries, from negative to positive (left to right). The taxonomic compositions corresponding to variable entries are shown for each of 80 equally spaced bins. Families belonging to the Order Rhizobiales are color-coded.}
\end{figure}

\begin{figure} [H]
    \centering
    \includegraphics[width=\textwidth]{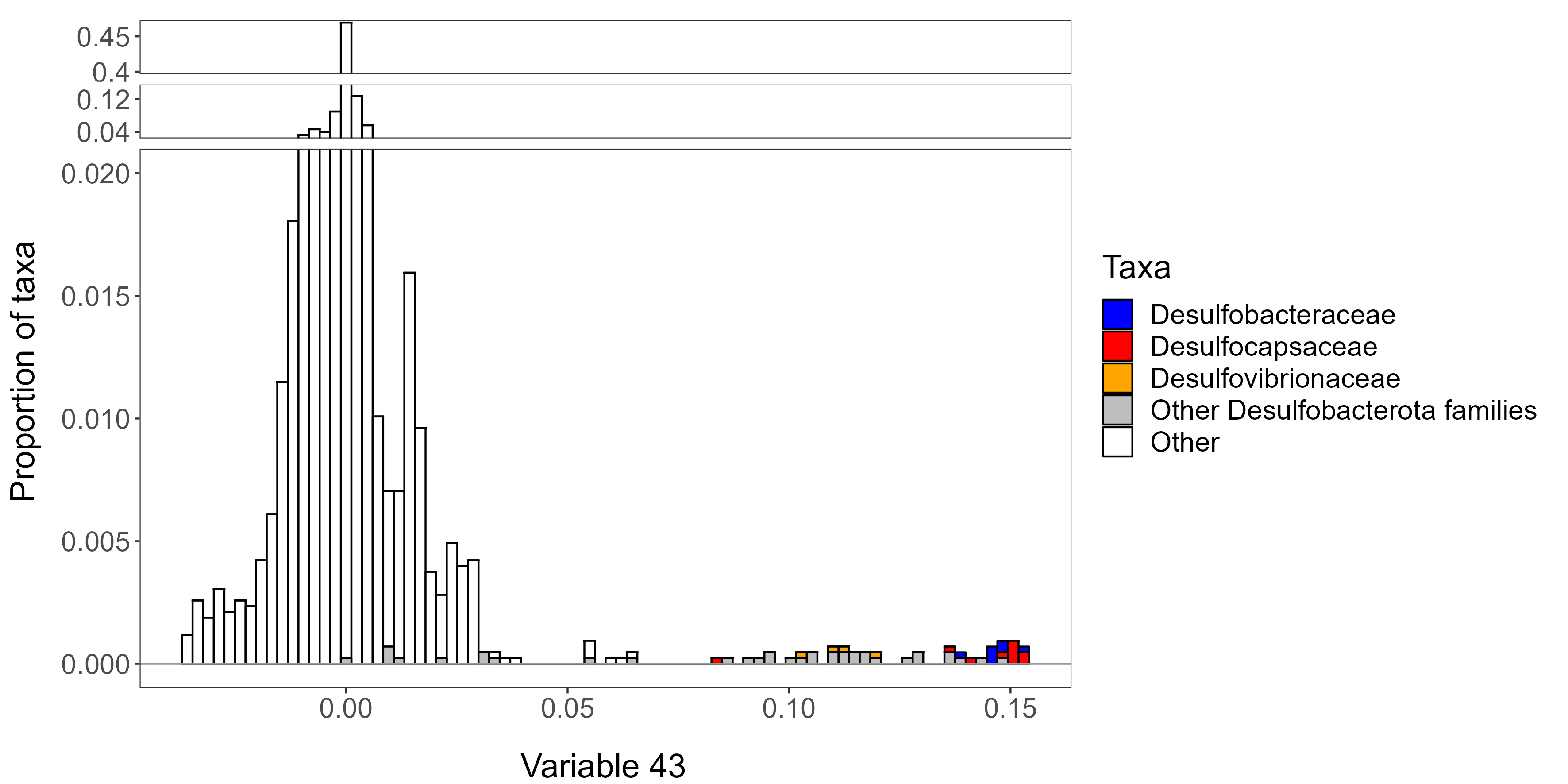}
    \caption{The ordering of taxa defined by variable 43 entries, from negative to positive (left to right). The taxonomic compositions corresponding to variable entries are shown for each of 80 equally spaced bins. Families belonging to the phylum Desulfobacterota are color-coded.}
\end{figure}

\begin{figure} [H]
    \centering
    \includegraphics[width=\textwidth]{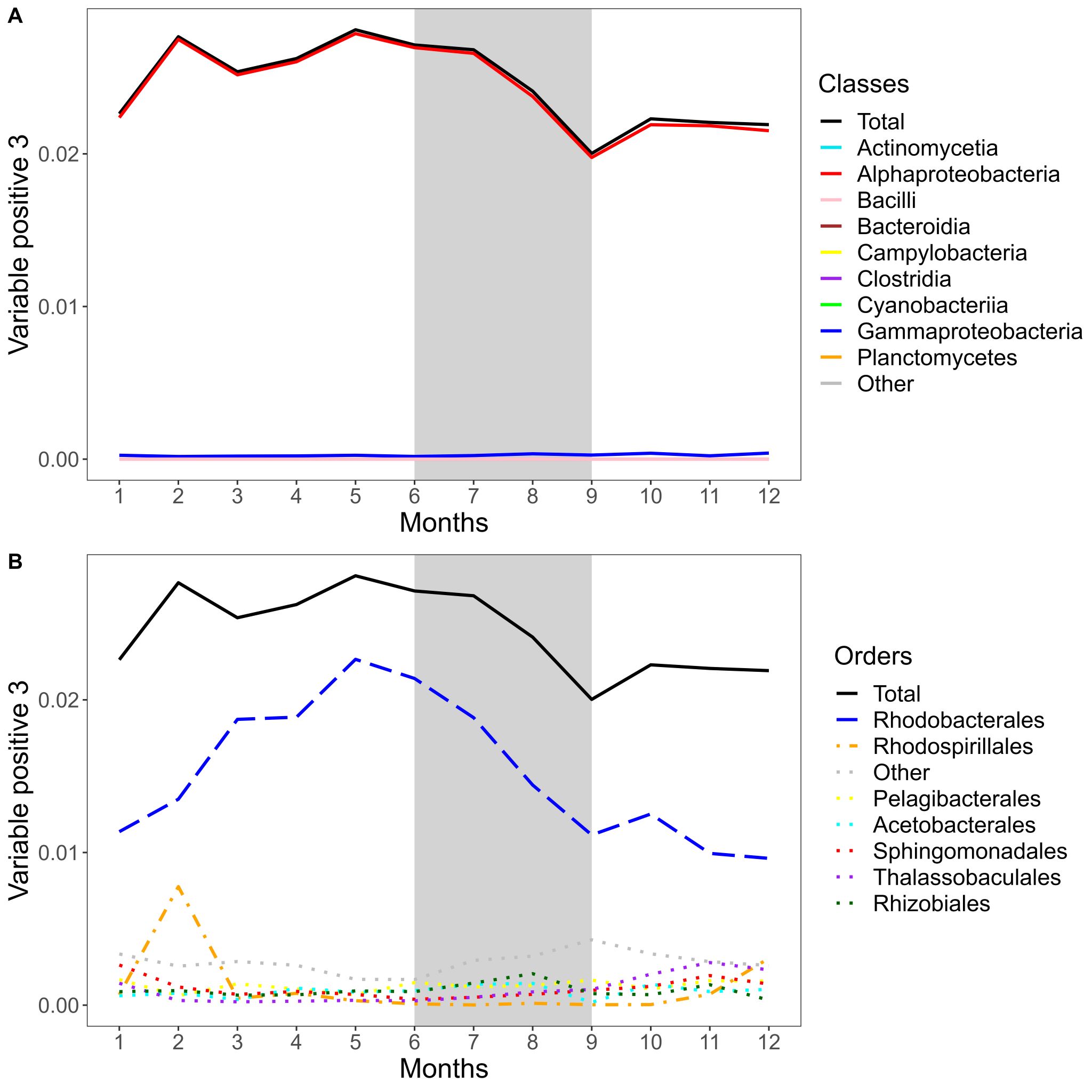}
    \caption{Abundance-weighted mean values of inferred ability of utilizing a variety of carbon sources over the yearly cycle. Summer months are indicated by a gray background. Taxonomic class (A) and taxonomic orders (B) are color-coded.}
\end{figure}

\begin{figure} [H]
    \centering
    \includegraphics[width=\textwidth]{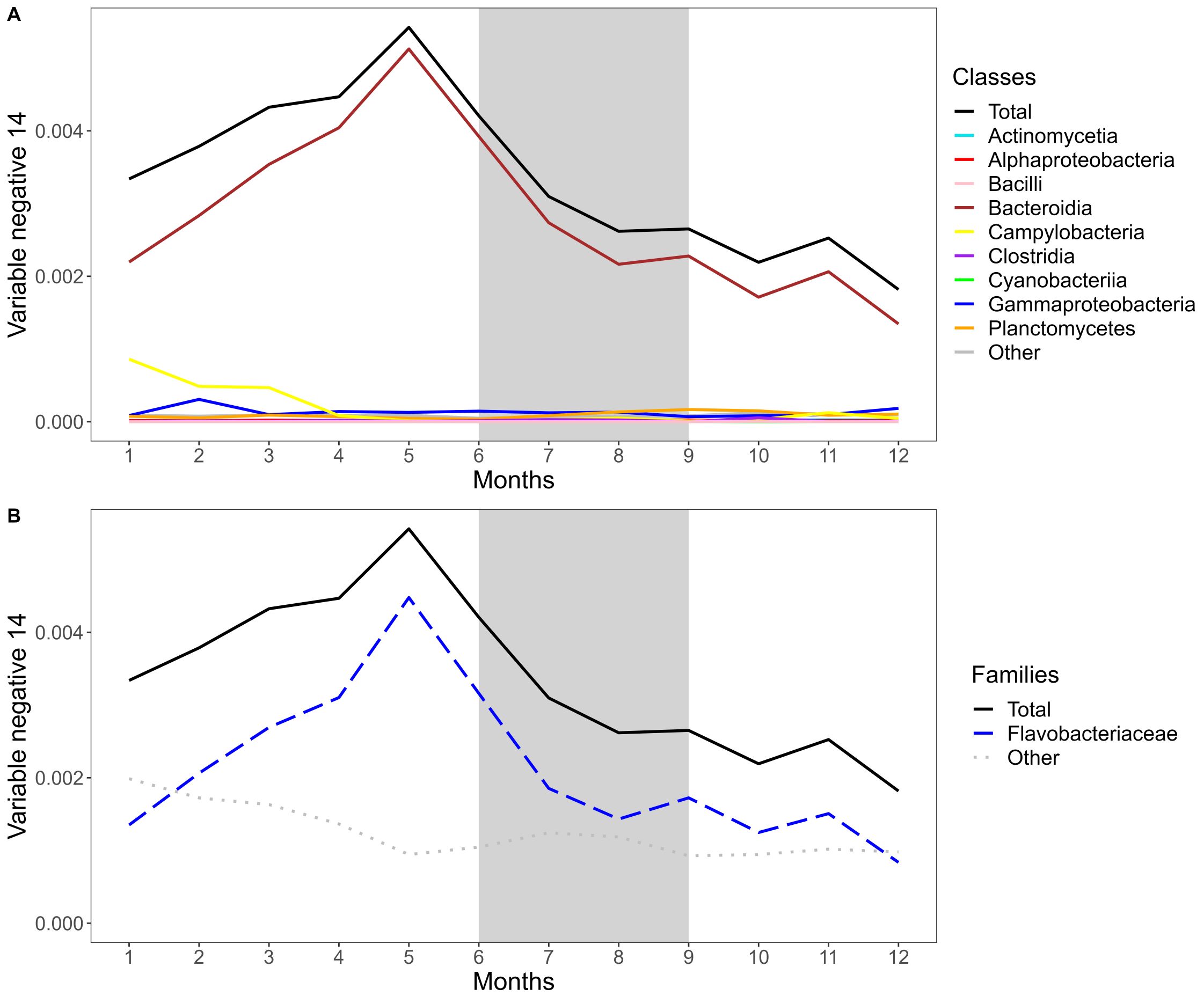}
    \caption{Abundance-weighted mean values of inferred ability of degrading complex polysaccharides over the yearly cycle. Summer months are indicated by a gray background. Taxonomic class (A) and taxonomic families (B) are color-coded.}
\end{figure}

\begin{figure} [H]
    \centering
    \includegraphics[width=\textwidth]{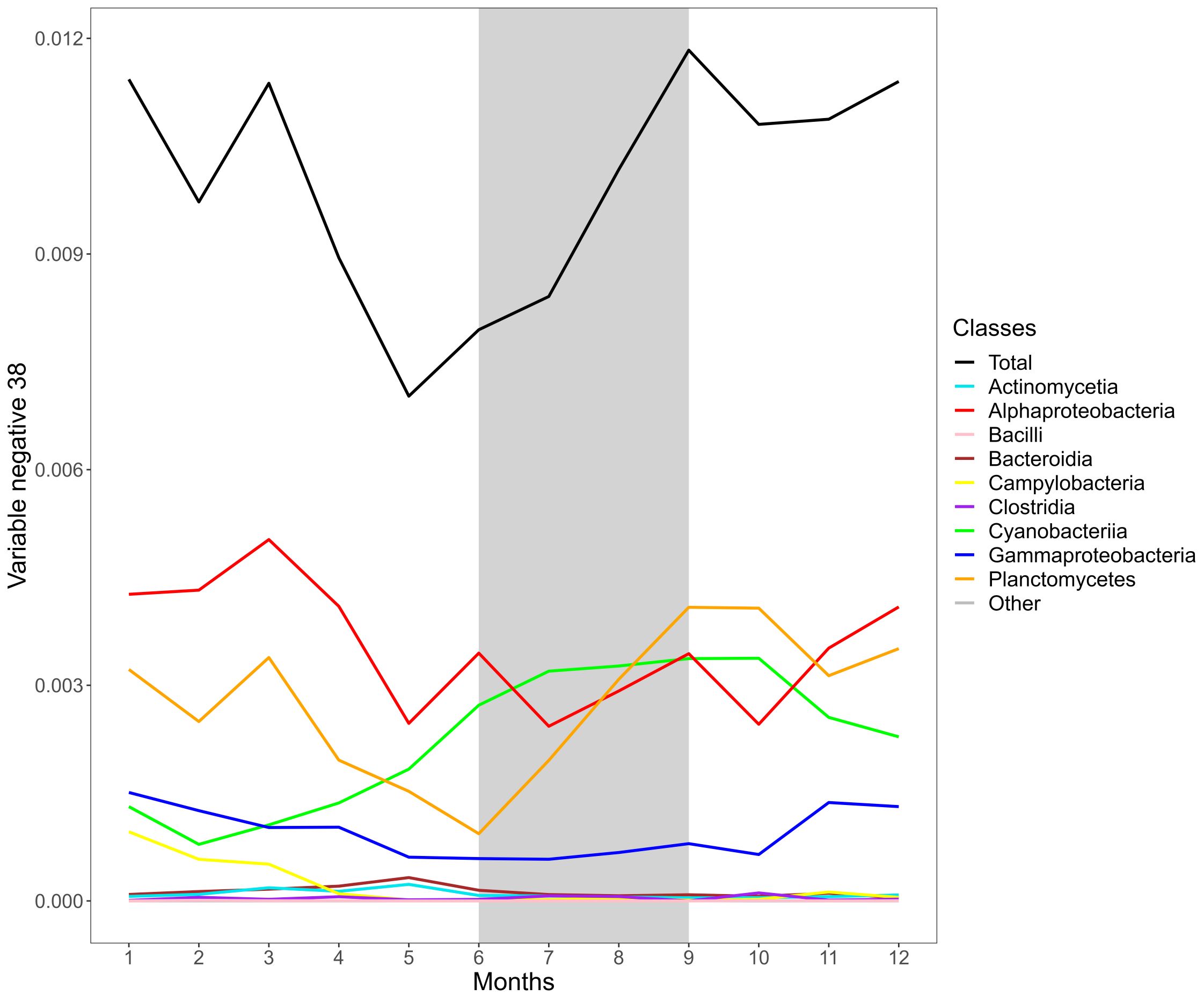}
    \caption{Abundance-weighted mean values of inferred ability of oxidizing methyl groups and C1 compounds over the yearly cycle. Summer months are indicated by a gray background. Taxonomic class is color-coded.}
\end{figure}

\begin{figure} [H]
    \centering
    \includegraphics[width=\textwidth]{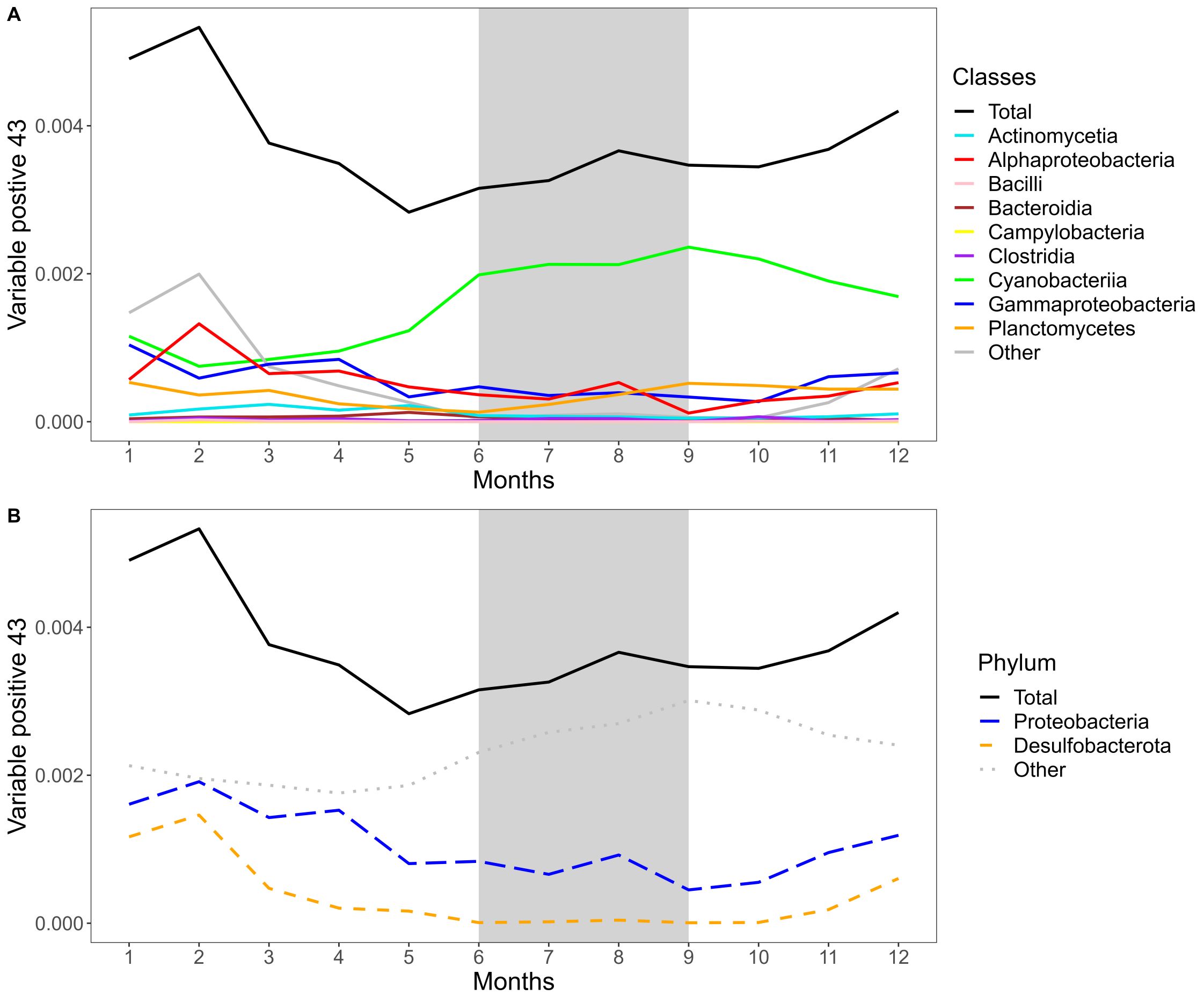}
    \caption{Abundance-weighted mean values of trait dominated by non-spore forming sulfate reducers over the yearly cycle. Summer months are indicated by a gray background. Taxonomic class is color-coded.}
\end{figure}

\bibliographystyle{unsrtnat}

\end{flushleft}